\documentclass[zpreprint,zbstepj]{./LaTeX/zeus/zeus_paper}
\usepackage[english]{babel}

\newcommand{\ZcoosysB}{%
The ZEUS coordinate system is a right-handed Cartesian system, with the $Z$
axis pointing in the proton beam direction, referred to as the ``forward
direction'', and the $X$ axis pointing left towards the centre of HERA.
The polar angle, $\theta$, is measured with respect to the proton beam
direction. The coordinate origin is at the nominal interaction point.\xspace}

\newcommand{\ZcoosysfnB}{\footnote{\ZcoosysB}}

\newcommand{\Zdetdesc}{%
A detailed description of the ZEUS detector can be found 
elsewhere~\cite{zeus:1993:bluebook}. A brief outline of the 
components most relevant for this analysis is given
below.\xspace}
\newcommand{\Zctddesc}[1]{%
Charged particles are tracked in the central tracking detector (CTD)~\citeCTD,
which operates in a magnetic field of $1.43\Tesla$ provided by a thin 
superconducting solenoid. The CTD consists of 72~cylindrical drift chamber 
layers, organised in nine~superlayers covering the polar-angle\ZcoosysfnB~region 
\mbox{$15^\circ<\theta<164^\circ$}. The relative transverse-momentum 
resolution for
full-length tracks is $\sigma(p_T)/p_T=0.0058p_T\oplus0.0065\oplus0.0014/p_T$,
with $p_T$ in $\Gev$. The position of the interaction vertex along the beam direction can be reconstructed from the CTD tracks with a resolution of about 1 cm in CC events.\xspace\\}
\newcommand{\Zcaldesc}{%
\\
The high-resolution uranium--scintillator calorimeter (CAL)~\citeCAL consists 
of three parts: the forward (FCAL), the barrel (BCAL) and the rear (RCAL)
calorimeter, covering 99.7\% of the solid angle around the nominal interaction point. 
Each part is subdivided transversely into towers and
longitudinally into one electromagnetic section (EMC) and either one (in RCAL)
or two (in BCAL and FCAL) hadronic sections (HAC). The smallest subdivision of
the calorimeter is called a cell.  The CAL relative energy resolutions, 
as measured under
test-beam conditions, are $\sigma(E)/E=0.18/\sqrt{E}$ for electrons and
$\sigma(E)/E=0.35/\sqrt{E}$ for hadrons, with $E$ in $\Gev$. The timing resolution of the CAL is better than 1 ns for energy deposits exceeding 4.5 \gev. The position of the interaction vertex along the beam direction can be reconstructed from the arrival time of energy deposits in FCAL. The resolution is about 9 cm for events with FCAL energy above 25 \gev, improving to about 7 cm for FCAL energy above 100 \gev.\xspace\\}

\chardef\usc=95
\chardef\til=126
\catcode`\@=11 
\DeclareRobustCommand\xdotspace{\futurelet\@let@token\@xdotspace}
\def\@xdotspace{%
  \ifx\@let@token.\else
  \ifx\@let@token\bgroup.\else
  \ifx\@let@token\egroup.\else
  \ifx\@let@token\/.\else
  \ifx\@let@token\ .\else
  \ifx\@let@token~.\else
  \ifx\@let@token!.\else
  \ifx\@let@token,.\else
  \ifx\@let@token:.\else
  \ifx\@let@token;.\else
  \ifx\@let@token?.\else
  \ifx\@let@token/.\else
  \ifx\@let@token'.\else
  \ifx\@let@token).\else
  \ifx\@let@token-.\else
  \ifx\@let@token\@xobeysp.\else
  \ifx\@let@token\space.\else
  \ifx\@let@token\@sptoken.\else
   .\space
   \fi\fi\fi\fi\fi\fi\fi\fi\fi\fi\fi\fi\fi\fi\fi\fi\fi\fi}
\catcode`\@=12 

\newcommand{\stru}[2]{%
   \relax\ifmmode\hbox{\vrule height#1 depth#2 width0pt}%
   \else\vrule height#1 depth#2 width0pt\fi}

\newcommand{\Ronum}[1]{\uppercase\expandafter{\romannumeral#1}}
\newcommand{\ronum}[1]{\expandafter{\romannumeral#1}}
\DeclareRobustCommand{\LaTeXZ}{%
  \LaTeX\kern-.05em4\kern-.1em
  {\raisebox{-0.2ex}{$\scriptstyle\text{ZEUS}$}}\xspace}

\DeclareMathAlphabet{\mathbf}{OT1}{cmr}{bx}{sl}
\newcommand{\eVdist}{\kern-0.06667em}

\newcommand{\Gev}{{\text{Ge}\eVdist\text{V\/}}}

\newcommand{\gev}{{\,\text{Ge}\eVdist\text{V\/}}}

\newcommand{\rad}{\,\text{rad}}

\newcommand{\Tesla}{\,\text{T}}

\newcommand{\slashfrac}[2]{%
  \raisebox{0.5ex}{\ensuremath #1}\kern-0.12em/\kern-0.08em
  \raisebox{-.8ex}{\ensuremath #2}}

\newcommand{\sqr}[3]{%
    {\vcenter{\hrule height.#3ex\hbox{\vrule width.#2ex height#1ex
     \kern#1ex\vrule width.#3ex}\hrule height.#2ex}}}

\newcommand{\widebar}[1]{%
   \mkern1.5mu\overline{\mkern-1.5mu#1\mkern-1.mu}\mkern1.mu}
\catcode`\@=11 
\newcommand{\parenbar}{\mathpalette\p@renb@r}
\def\p@renb@r#1#2{\vbox{%
  \ifx#1\scriptscriptstyle \dimen@.7em\dimen@ii.2em\else
  \ifx#1\scriptstyle \dimen@.8em\dimen@ii.25em\else
  \dimen@1em\dimen@ii.4em\fi\fi \offinterlineskip
  \ialign{\hfill##\hfill\cr
    \vbox{\hrule width\dimen@ii}\cr
    \noalign{\vskip-.3ex}%
    \hbox to\dimen@{$\mathchar300\hfil\mathchar301$}\cr
    \noalign{\vskip-.3ex}%
    $#1#2$\cr}}}
\catcode`\@=12 

\newcommand{\nubar}{\widebar{\nu}}

\newcommand{\MSbar}{\hbox{$\overline{\rm MS}$}\xspace}

\newcommand{\IP}{{\rm I$\kern-0.01667em$P}\xspace}

\mathchardef\qsm=63
\mathchardef\pls=43
\mathchardef\mns=512
\mathchardef\plm=518
\mathchardef\eql=61
\mathchardef\smallleft=300
\mathchardef\smallright=301
\mathchardef\les=316
\mathchardef\gre=318
\mathchardef\leq=532
\mathchardef\grq=533

\catcode`\@=11 
\newcounter{pict@width}
\newcounter{pict@height}
\newlength{\pict@scale}
\newcommand{\psfigadd}[4]{%
\setcounter{pict@width}{1*\ratio{#2+\pict@scale/2}{\pict@scale}}
\setcounter{pict@height}{1*\ratio{#3+\pict@scale/2}{\pict@scale}}
\setlength{\unitlength}{\pict@scale}
\hbox to #2{\hspace{-\fill}\begin{picture}(\thepict@width,\thepict@height)
\put(0,0){\psfig{figure=#1,width=#2,height=#3,clip=}}
\SetScale{0.283466457}
\SetWidth{1.763889}
{#4}
\end{picture}}
}
\newcounter{pict@widthfst}
\newcounter{pict@widthscd}
\newcounter{pict@widthtot}
\newcommand{\psfigaddtwo}[7]{%
\setcounter{pict@widthfst}{1*\ratio{#2+\pict@scale/2}{\pict@scale}}
\setcounter{pict@widthscd}{1*\ratio{#2+#4+\pict@scale/2}{\pict@scale}}
\setcounter{pict@widthtot}{1*\ratio{#2+#4+#6+\pict@scale/2}{\pict@scale}}
\setcounter{pict@height}{1*\ratio{#3+\pict@scale/2}{\pict@scale}}
\setlength{\unitlength}{\pict@scale}
\hbox{\hspace{-\fill}\begin{picture}(\thepict@widthtot,\thepict@height)
\put(0,0){\psfig{figure=#1,width=#2,height=#3,clip=}}
\put(\thepict@widthscd,0){\psfig{figure=#5,width=#6,height=#3,clip=}}
\SetScale{0.283466457}
\SetWidth{1.763889}
{#7}
\end{picture}}
}
\newcommand{\psfigror}[4]{%
\setcounter{pict@width}{1*\ratio{#2+\pict@scale/2}{\pict@scale}}
\setcounter{pict@height}{1*\ratio{#3+\pict@scale/2}{\pict@scale}}
\setlength{\unitlength}{\pict@scale}
\hbox{\begin{picture}(\thepict@width,\thepict@height)
\put(0,\thepict@height){\psfig{figure=#1,width=#3,height=#2,clip=,angle=270}}
\SetScale{0.283466457}
\SetWidth{1.763889}
{#4}
\end{picture}}
}
\newcommand{\psfigrol}[4]{%
\setcounter{pict@width}{1*\ratio{#2+\pict@scale/2}{\pict@scale}}
\setcounter{pict@height}{1*\ratio{#3+\pict@scale/2}{\pict@scale}}
\setlength{\unitlength}{\pict@scale}
\hbox{\begin{picture}(\thepict@width,\thepict@height)
\put(0,0){\psfig{figure=#1,width=#3,height=#2,clip=,angle=90}}
\SetScale{0.283466457}
\SetWidth{1.763889}
{#4}
\end{picture}}
}
\catcode`\@=12 
\newlength\listtextwidth

\catcode`\@=11 
\newlength{\@tabfninsert}
\newlength{\@tabfnwidth}
\newcommand{\tabfootnote}[2]{%
  \setlength{\@tabfninsert}{0.8em}
  \setlength{\@tabfnwidth}{\textwidth}
  \addtolength{\@tabfnwidth}{-\@tabfninsert}
  \addtolength{\@tabfnwidth}{-0.4em}
  \noindent\makebox[\@tabfninsert][r]{\footnotesize$^{#1}$\hfil}\hfill%
  \parbox[t]{\@tabfnwidth}{\footnotesize #2\hfill}}
\catcode`\@=12 

\newcommand{\PTM}       {P_{T,{\rm miss}}}

\def\citeCTD{{\cite{%
nim:a279:290,*npps:b32:181,*nim:a338:254%
}}\xspace}
\def\citeCAL{{\cite{%
nim:a309:77,*nim:a309:101,*nim:a321:356,*nim:a336:23%
}}\xspace}

\begin{document}
\title{
Measurement of high-\boldmath${Q^2}$ charged current\\
cross sections in \boldmath${e^+p}$ deep inelastic  \\
scattering at HERA
}                                                       
                    
\author{ZEUS Collaboration}

\date{}

\prepnum{{DESY--03--093}}

\abstract{
Cross sections for $e^+p$ charged current deep inelastic scattering at a
centre-of-mass energy of $318\gev$ have been determined with 
an integrated luminosity of $60.9~\rm{pb^{-1}}$ 
collected with the ZEUS detector at HERA. 
The differential cross sections $d\sigma/d Q^2$, $d\sigma/d x$ and $d\sigma/d y$ for \mbox{$Q^{2}>200\gev^2$} are presented. In addition, $d^2 \sigma/d x d Q^2$ has been measured in the kinematic range \mbox{$280\gev^2 <Q^2<17\,000 \gev^2$} and \mbox{$0.008<x<0.42$}. The predictions of the Standard Model agree well with the measured cross sections. The mass of the $W$ boson propagator is determined to be \mbox{$M_W=78.9\pm 2.0\,{\rm (stat.)}\,\pm 1.8\,{\rm(syst.)}\,^{+2.0}_{-1.8}\,{\rm(PDF)}\,\gev$} from a fit to $d \sigma/ d Q^2$. The chiral structure of the Standard Model is also investigated in terms of the $(1-y)^2$ dependence of the double-differential cross section. The structure-function $F_{2}^{\rm{CC}}$ has been extracted by combining the measurements presented here with previous ZEUS results from $e^- p$ scattering, extending the measurement obtained in a neutrino-nucleus scattering experiment to a significantly higher $Q^{2}$ region.}

\makezeustitle

\def\3{\ss}                                                                                        
\pagenumbering{Roman}                                                                              
                                                   %
\begin{center}                                                                                     
{                      \Large  The ZEUS Collaboration              }                               
\end{center}                                                                                       
  S.~Chekanov,                                                                                     
  M.~Derrick,                                                                                      
  D.~Krakauer,                                                                                     
  J.H.~Loizides$^{   1}$,                                                                          
  S.~Magill,                                                                                       
  B.~Musgrave,                                                                                     
  J.~Repond,                                                                                       
  R.~Yoshida\\                                                                                     
 {\it Argonne National Laboratory, Argonne, Illinois 60439-4815}~$^{n}$                            
\par \filbreak                                                                                     
  M.C.K.~Mattingly \\                                                                              
 {\it Andrews University, Berrien Springs, Michigan 49104-0380}                                    
\par \filbreak                                                                                     
  P.~Antonioli,                                                                                    
  G.~Bari,                                                                                         
  M.~Basile,                                                                                       
  L.~Bellagamba,                                                                                   
  D.~Boscherini,                                                                                   
  A.~Bruni,                                                                                        
  G.~Bruni,                                                                                        
  G.~Cara~Romeo,                                                                                   
  L.~Cifarelli,                                                                                    
  F.~Cindolo,                                                                                      
  A.~Contin,                                                                                       
  M.~Corradi,                                                                                      
  S.~De~Pasquale,                                                                                  
  P.~Giusti,                                                                                       
  G.~Iacobucci,                                                                                    
  A.~Margotti,                                                                                     
  R.~Nania,                                                                                        
  F.~Palmonari,                                                                                    
  A.~Pesci,                                                                                        
  G.~Sartorelli,                                                                                   
  A.~Zichichi  \\                                                                                  
  {\it University and INFN Bologna, Bologna, Italy}~$^{e}$
% ...
\newpage                                                                                           
$^{\    1}$ also affiliated with University College London \\                                      
$^{\    2}$ on leave of absence at University of                                                   
Erlangen-N\"urnberg, Germany\\                                                                     
% ...
$^{  17}$ now at Univ. of Oxford, Oxford/UK \\                                                     
% ...
                                                           %
                                                           %
\newpage   
                                                           %
                                                           %
\begin{tabular}[h]{rp{14cm}}                                                                       
$^{a}$ &  supported by the Natural Sciences and Engineering Research                               
          Council of Canada (NSERC) \\                                                             
$^{b}$ &  supported by the German Federal Ministry for Education and                               
          Research (BMBF), under contract numbers HZ1GUA 2, HZ1GUB 0, HZ1PDA 5, HZ1VFA 5\\         
$^{c}$ &  supported by the MINERVA Gesellschaft f\"ur Forschung GmbH, the                          
          Israel Science Foundation, the U.S.-Israel Binational Science                            
          Foundation and the Benozyio Center                                                       
          for High Energy Physics\\                                                                
$^{d}$ &  supported by the German-Israeli Foundation and the Israel Science                        
          Foundation\\                                                                             
$^{e}$ &  supported by the Italian National Institute for Nuclear Physics (INFN) \\                
$^{f}$ &  supported by the Japanese Ministry of Education, Culture,                                
          Sports, Science and Technology (MEXT) and its grants for                                 
          Scientific Research\\                                                                    
$^{g}$ &  supported by the Korean Ministry of Education and Korea Science                          
          and Engineering Foundation\\                                                             
$^{h}$ &  supported by the Netherlands Foundation for Research on Matter (FOM)\\                   
$^{i}$ &  supported by the Polish State Committee for Scientific Research,                         
          grant no. 620/E-77/SPUB-M/DESY/P-03/DZ 247/2000-2002\\                                   
$^{j}$ &  partially supported by the German Federal Ministry for Education                         
          and Research (BMBF)\\                                                                    
$^{k}$ &  supported by the Fund for Fundamental Research of Russian Ministry                       
          for Science and Edu\-cation and by the German Federal Ministry for                       
          Education and Research (BMBF)\\                                                          
$^{l}$ &  supported by the Spanish Ministry of Education and Science                               
          through funds provided by CICYT\\                                                        
$^{m}$ &  supported by the Particle Physics and Astronomy Research Council, UK\\                   
$^{n}$ &  supported by the US Department of Energy\\                                               
$^{o}$ &  supported by the US National Science Foundation\\                                        
$^{p}$ &  supported by the Polish State Committee for Scientific Research,                         
          grant no. 112/E-356/SPUB-M/DESY/P-03/DZ 301/2000-2002, 2 P03B 13922\\                    
$^{q}$ &  supported by the Polish State Committee for Scientific Research,                         
          grant no. 115/E-343/SPUB-M/DESY/P-03/DZ 121/2001-2002, 2 P03B 07022\\                    
\end{tabular}                                                                                      
                                                           %
                                                           %

\pagenumbering{arabic} 
\pagestyle{plain}
\section{\bf Introduction}
\label{s:introduction}
Measurements of deep inelastic scattering (DIS) of leptons on nuclei
have been vital in the development of our understanding
of the structure of nucleons.
In the Standard Model (SM),
charged current (CC) DIS is mediated by the exchange of 
the $W$ boson.
In contrast to neutral current (NC) interactions, where all quark and
antiquark flavours participate, only down-type quarks
and up-type antiquarks 
participate at leading order in the $e^+p$ CC DIS reaction.
Thus, such interactions are a powerful tool for flavour-specific 
investigations of the parton distribution functions (PDFs). 
Since only left-handed quarks and right-handed antiquarks contribute
to CC DIS, the distribution of the positron-quark 
centre-of-mass scattering angle, $\theta^*$, is a sensitive probe of the
chiral structure of the weak interaction.

Measurements of the CC DIS cross sections at HERA have been
reported previously by the H1\mcite{pl:b324:241,zfp:c67:565,pl:b379:319,hep-ex-0304003}
and ZEUS\cite{prl:75:1006,zfp:c72:47} collaborations. 
These data extended the kinematic region covered by fixed-target 
neutrino-nucleus scattering experiments\mcite{zfp:c25:29,zfp:c49:187,zfp:c53:51,zfp:c62:575}
by about two orders of magnitude towards larger $Q^2$, the negative square of the four-momentum transfer. 
In addition, the double-differential cross section, 
$d^2\sigma / dxdQ^2$, where $x$ is the Bjorken scaling variable, was measured for the first time 
by the HERA collider experiments in $e^+ p$ scattering\cite{epj:c13:609,epj:c12:411} 
and more recently in $e^- p$ scattering\cite{epj:c19:269,pl:b539:197}.

This paper presents measurements of the
$e^+p$ CC DIS single-differential cross sections $ d \sigma/ d  Q^2$, 
$ d \sigma/ d  x$ and  $ d \sigma/ d  y$,
and of $d^2 \sigma / dxdQ^2$ and their comparison with the SM.
The measurements are based on $60.9~\rm{pb^{-1}}$ of data collected
during the running periods in 1999 and 2000 when HERA 
collided $27.5\gev$ positrons with $920\gev$ protons, 
yielding a centre-of-mass energy ($\sqrt{s}$) of $318\gev$.  
The previous
ZEUS measurement~\cite{epj:c12:411} at $\sqrt{s}=300\gev$ was based on $47.7~\rm{pb^{-1}}$. 
The mass of the $W$ boson is determined from the measured cross sections and the helicity 
structure of the Standard Model is investigated.
The structure-function $F_{2}^{\rm{CC}}$ is extracted 
and compared to a fixed-target result~\cite{prl:86:2742}. 

\section{\bf Standard Model prediction}
\label{s:KinematicsSM}
The CC DIS differential cross section, $d^2\sigma^{CC}_{\rm Born}/dxdQ^2$, 
for the reaction $e^{+} p \rightarrow \nubar_e X$ can be written
at leading order in the electroweak interaction, for longitudinally unpolarised beams as
~\cite{ijmp:a13:3385}: 
\begin{eqnarray}
\frac{d^{2}\sigma^{\rm{CC}}_{\rm{Born}}(e^{+}p)}{dxdQ^{2}} & = & \lefteqn{\frac{G^{2}_{\rm{F}}}{4\pi x}
\frac{M_{W}^{4}}{(Q^{2}+M_{W}^{2})^{2}}\times} \nonumber \\& & [Y_{+}F_{2,e^{+}p}^{\rm{CC}}(x,Q^{2})-y^{2}F_{{\rm{L}},e^{+}p}^
{\rm{CC}}(x,Q^{2}) - Y_{-}xF_{3,e^{+}p}^{\rm{CC}}(x,Q^{2})].
\label{e:Bornpos}
\end{eqnarray}
In this equation, $G_{\rm{F}}$ is the Fermi constant, $M_W$ is the mass of the $W$ boson, $x$ is the Bjorken scaling variable, $y=Q^2 /xs$ (neglecting the masses of the incoming particles) and $Y_{\pm}=1\pm (1-y)^2$. The centre-of-mass energy in the positron-proton collisions is given by $\sqrt{s}=2\sqrt{E_{e}E_{p}}$, where $E_{e}$ and $E_{p}$ are the positron and proton beam energies, respectively. The inelasticity, $y$, is related to $\theta^*$ by $y=(1-\cos\theta^*)/2$. The structure-functions $F_{2,e^{+}p}^{\rm{CC}}$ and $xF_{3,e^{+}p}^{\rm{CC}}$, at leading order in QCD, may be written in terms of sums and differences of quark and antiquark PDFs of the proton as follows:
\begin{equation}
F_{2,e^{+}p}^{CC} = x[d(x,Q^{2})+s(x,Q^{2})+\bar{u}(x,Q^{2})+\bar{c}(x,Q^{2})],\label{e:f2pos}
\end{equation}
\begin{equation}
xF_{3,e^{+}p}^{CC} = x[d(x,Q^{2})+s(x,Q^{2})-\bar{u}(x,Q^{2})-\bar{c}(x,Q^{2})],\label{e:xf3pos}
\end{equation}
where, for example, the PDF $d(x,Q^{2})$ gives the number density of down quarks
with momentum-fraction $x$ at a given $Q^2$. The longitudinal structure
function, $F_{{\rm{L}},e^{+}p}^{CC}$, is zero at leading order in QCD. At
next-to-leading order (NLO), $F_{{\rm{L}},e^{+}p}^{CC}$ is non-zero but
gives a negligible contribution to the cross section except at values of $y$
close to 1, where the contribution can be as large as 10\%. Since the top-quark mass is large and the off-diagonal elements of the CKM matrix are small~\cite{epj:c15:1}, the contribution from third-generation quarks may be ignored~\cite{katz:2000:hera}. All cross-section calculations presented in this paper were performed at NLO in the strong coupling constant.
\section{\bf The ZEUS experiment}
\label{s:detector}
\Zdetdesc

\Zctddesc

\Zcaldesc

An iron structure that serves as a flux-return path for the magnetic field
of the solenoid surrounds the CAL and is instrumented as a backing
calorimeter (BAC)~\cite{nim:a313:126}. Muon chambers in the forward, barrel and rear~\cite{nim:a333:342} regions are used in this analysis to veto background events induced by cosmic-ray or beam-halo muons.

The luminosity was measured using the Bethe-Heitler reaction $ep
\rightarrow e \gamma p$. 
The photons were detected by the luminosity monitor~\cite{desy-92-066,*zfp:c63:391,*acpp:b32:2025}, a lead--scintillator calorimeter placed in the HERA tunnel 107 m from the interaction point in the positron beam direction. Events corresponding to an integrated luminosity of $60.9\pm1.5~\rm{pb}^{-1}$ were used in this analysis.

\section{Monte Carlo simulation}
\label{s:MCsimulation}
Monte Carlo (MC) simulations were used to determine the efficiency for 
selecting events and the accuracy of kinematic 
reconstruction, to estimate the $ep$ background rates and to extract
cross sections for the full kinematic region.
A sufficient number of
events was generated to ensure that the statistical uncertainties arising from the MC simulation were negligible compared to those of the data.
The MC samples were normalised to the total
integrated luminosity of the data.

Charged current DIS events, including electroweak radiative effects, were 
simulated using the {\sc heracles} 4.6.1~\cite{cpc:69:155,*spi:www:heracles} 
program with the
{\sc djangoh} 1.1~\cite{spi:www:djangoh11} interface to the MC generators that provide the hadronisation. Initial-state radiation, vertex and 
propagator corrections and two-boson exchange are included in {\sc heracles}.
The mass of the $W$ boson was calculated using the PDG~\cite{epj:c15:1} values for the fine structure constant, the Fermi constant, the masses of the $Z$ boson and the top quark, and with the Higgs-boson mass set to $100~\gev$.
The events were generated using the CTEQ5D~\cite{epj:c12:375} PDFs.
The colour-dipole model of {\sc ariadne} 4.10~\cite{cpc:71:15}
was used to simulate $\mathcal{O}(\alpha_{S})$ plus leading logarithmic corrections to the result of the quark-parton model. As a systematic check, the {\sc meps} model of
{\sc lepto} 6.5~\cite{cpc:101:108} was used.
Both programs use the Lund string model of {\sc jetset} 7.4~\cite{cpc:39:347,*cpc:43:367,*cpc:82:74}
for the hadronisation.
A set of NC events generated with {\sc djangoh} was used to estimate
the NC contamination in the CC sample.
Photoproduction background was estimated using events 
simulated with {\sc herwig} 5.9~\cite{cpc:67:465}.
The background from 
$W$ production was estimated
using the {\sc epvec}~\cite{np:b375:3} generator, and the background from
production of charged-lepton pairs was generated with the {\sc lpair}~\cite{proc:hera:1991:1478} and {\sc grape}~\cite{cpc:136:126} programs.

The ZEUS detector response was simulated with a program based on 
{\sc geant}~3.13~\cite{tech:cern-dd-ee-84-1}.  The simulated events were subjected 
to the same trigger requirements as 
the data, and processed by the same reconstruction programs.
\section{Reconstruction of kinematic variables}
\label{s:reconstruction}
The principal signature of CC DIS at HERA
is the presence of a large
missing transverse momentum, $\PTM$,
arising from the energetic final-state neutrino which escapes detection. 
The quantity $\PTM$ was calculated from
\begin{equation}
\PTM^2  =  P_X^2 + P_Y^2 = 
  \left( \sum\limits_{i} E_i \sin \theta_i \cos \phi_i \right)^2
+ \left( \sum\limits_{i} E_i \sin \theta_i \sin \phi_i \right)^2, \nonumber
  \label{eq:pt}
\end{equation}
where the sums run over all CAL energy deposits, $E_i$ 
(uncorrected in the trigger, but corrected~\cite{epj:c11:427} for energy loss 
in inactive material and other effects in the offline analysis), and
$\theta_i$ and $\phi_i$ are the polar and azimuthal angles
of the calorimeter deposits as viewed from the interaction vertex.
The polar angle of the hadronic system, $\gamma_h$, is defined by
$\cos\gamma_h = (\PTM^2 - \delta^2)/(\PTM^2 + \delta^2)$,
where
$\delta = \sum\limits_{i} E_i ( 1 - \cos \theta_{i} ) 
= \sum\limits_{i} (E-P_Z)_{i}$.
In the naive quark-parton model,
$\gamma_h$ is the angle through which the struck quark is scattered.
Finally, the total transverse energy,
$E_T$, is given by
$E_T    = \sum\limits_{i} E_i \sin \theta_i$. 

 The kinematic variables were reconstructed
using the Jacquet-Blondel method \cite{proc:epfacility:1979:391}.
The estimators of $y$, $Q^2$ and $x$ are:
$y_{\rm{JB}} = \delta/(2E_e)$,
$Q^2_{\rm{JB}} = \PTM^2/(1-y_{\rm{JB}})$, and
$x_{\rm{JB}} = Q^2_{\rm{JB}}/(sy_{\rm{JB}})$.

The resolution in $Q^2$ 
is about $20\%$. The resolution in $x$ improves from about $20$\%
at $x=0.01$ to about $5\%$ at $x=0.5$.
The resolution in $y$ ranges from about $14\%$ at $y=0.05$ to about 
$8\%$ at $y=0.83$.

\section{Event selection}
\label{s:EvSel}
Charged current DIS candidates were selected by requiring a large $\PTM$ and a 
reconstructed event vertex consistent with an $ep$ interaction.
The main sources of background come from NC scattering and high-$E_T$
photoproduction in which the finite energy resolution of the CAL
or energy that escapes detection
can lead to significant measured missing transverse momentum.  
Non-$ep$ events such as beam-gas interactions,
beam-halo muons or cosmic rays can also cause substantial 
imbalance in the measured transverse
momentum and constitute additional sources of background.
The selection criteria described below were imposed
to separate CC events from all backgrounds.

When the current jet lies in the central region of the detector, i.e. $\gamma_h$ is large, 
tracks in the CTD can be used to reconstruct 
the event vertex, strongly suppressing non-$ep$ backgrounds.
For CC events with small $\gamma_h$, the charged particles of
the hadronic final state 
are often outside the acceptance of the CTD.
Such events populate the high-$x$ region of the kinematic plane.
The events were classified first according to $\gamma_0$, the value of
$\gamma_h$ measured with respect to the nominal interaction point.
Subsequently, the kinematic quantities were recalculated using the
$Z$-coordinate of the event vertex ($Z_{\rm VTX}$) determined
from either CTD tracks, for events with large $\gamma_0$, or the calorimeter-timing 
information for events in which $\gamma_0$ is small. The selection procedures for events 
with large and small $\gamma_0$ are described in Sections \ref{ss:StanEvSel} and \ref{ss:LowGEvSel}, respectively.

\subsection{Trigger selection}
\label{ss:Trigger}
ZEUS has a three-level trigger system~\cite{zeus:1993:bluebook,smith:1992}. 
At the first
level only coarse calorimeter and tracking information is available.
Events were selected using criteria based on the energy, transverse 
energy and missing transverse momentum measured in the calorimeter. 
Generally, events were triggered with low thresholds on these quantities if a
coincidence with CTD tracks from the event vertex occurred, while higher 
thresholds were required for events with no CTD 
tracks. Typical threshold values were $5~\gev$~($8~\gev$)~in missing 
transverse momentum, or $11.5~\gev$~($21~\gev$)~in transverse energy for events with (without) CTD tracks.

At the second level, timing information from the calorimeter was used to reject events inconsistent with the bunch-crossing time. In addition, the topology of the CAL energy deposits was used to reject background events. 
In particular, a tighter cut of $6~\gev$~($9~\gev$~for events without CTD
tracks) was made on missing transverse momentum, since the resolution in this variable is
better at the second level than at the first level.

At the third level, full track reconstruction and vertex finding were performed
and used to reject candidate events with a vertex inconsistent with an $ep$
interaction. Cuts were applied to calorimeter quantities and reconstructed
tracks to further reduce beam-gas contamination.

The efficiency of the full trigger chain for CC DIS events, in the kinematic region of the cross-section measurements, was determined using MC simulation to be 96\%, and not below 78\% in any cross-section bin.

\subsection{Offline selection based on a CTD vertex}
\label{ss:StanEvSel}
Events with $\gamma_{0}>0.4$ rad
were required to have a vertex reconstructed from CTD tracks. Additional requirements for event
selection are described below.

\begin{itemize}
  \item{Primary vertex: events were required to satisfy $| Z_{\rm VTX} | < 50$~cm. 
        The primary vertex reconstructed from the CTD
        tracks was required to be within the range consistent with the $ep$
        interaction region. The transverse sizes of the beams were smaller than 
        the CTD vertex resolution in X and Y. Furthermore, the average transverse 
        displacement of the vertex from its nominal position, as measured by
        the CTD, was also smaller than the resolution. Therefore, the X- and
        Y-vertex positions were set to zero, their nominal positions;}
  \item{Missing transverse momentum: events were required to satisfy $\PTM > 12~\gev$ 
        and $\PTM ' > 10~\gev$ where $\PTM '$ is the missing
        transverse momentum calculated excluding the FCAL towers closest to the
        beam hole. The $\PTM '$ cut strongly suppresses beam-gas events while
        maintaining high efficiency for CC events;}
  \item{Tracking requirement: tracks associated with the event vertex with
        transverse momentum in excess of $0.2\gev$ and a polar
        angle in the range $15^\circ$ to $164^\circ$ were defined as ``good''
        tracks. In order to remove
        beam-gas background, at least one such track was required and a cut was 
        also applied in two dimensions on the
        number of good tracks versus the total number of tracks;}
  \item{Rejection of photoproduction:
  \begin{itemize} 
  \item{$\PTM/E_T > 0.4$ was required for events
        with $\PTM < 30\gev$. This cut was raised to $\PTM/E_T > 0.55$ for
        events with $\PTM < 20\gev$. This selected events with a collimated energy flow,
        as expected from a single scattered quark;}

  \item{for events with $\PTM < 20\gev$, the number of good tracks within an
        azimuthal angle of $0.5 \rad$ around the direction of the $\PTM$ was
        counted ($N_{+}$) as well as the number of tracks opposite to the $\PTM$ 
        direction ($N_{-}$). The event was rejected if the number of tracks in
        the $\PTM$ direction was $\geq 2$, or if the asymmetry defined as 
        $(N_{-}-N_{+})/(N_{-}+N_{+})$ was less than 0.7;}
 
  \item{for charged current events, there is a correlation between the direction
        of the $\PTM$ vector calculated using CTD tracks and that obtained using
        the CAL. The difference between these quantities was required to be less
        than 0.5~radians for $\PTM <20\gev$ and less than 2.0~radians for $\PTM \geq 20\gev$;}
  \end{itemize}}
  \item{Rejection of NC DIS: NC DIS events in which the scattered positron or
        jet energies are poorly measured can have a considerable apparent
        missing transverse momentum. To identify such events, a search for
        candidate positrons was made using isolated electromagnetic clusters in
        the CAL~\cite{nim:a365:508,*nim:a391:360} for events with $\PTM <
        30\gev$. Candidate positron clusters within the CTD acceptance were
        required to have an energy above $4\gev$ and a matching track with
        momentum larger than 25\% of the cluster energy. Clusters with $\theta >
        164^\circ$ were required to have a transverse momentum exceeding
        $2\gev$. Events with a candidate positron satisfying the above criteria
        and $\delta > 30\gev$ were rejected, since for fully contained NC
        events, $\delta$ peaks at $2E_e = 55\gev$;}
  \item{Rejection of non-$ep$ background: muon-finding
        algorithms~\cite{thesis:kruse:1999} based on CAL energy deposits or
        muon-chamber signals were used to reject events produced by cosmic rays
        or muons in the beam halo.}
\end{itemize}

\subsection{Offline selection based on a CAL vertex}
\label{ss:LowGEvSel}
For events with $\gamma_0<0.4$ rad a vertex reconstructed using calorimeter timing information was required. Additional requirements for event selection are described below.

\begin{itemize}
  \item{Primary vertex: events were required to satisfy $| Z_{\rm VTX} | < 50$~cm. 
        $Z_{\rm VTX}$ was reconstructed from the
        measured arrival time of energy deposits in FCAL~\cite{pl:b316:412}. The
        $X$- and $Y$-vertex positions were set to zero;}
  \item{Missing transverse momentum: events were required to satisfy 
        $\PTM > 14\gev$ and $\PTM ' > 12\gev$. 
        The conditions on missing
        transverse momentum were tightened, compensating for the relaxation of
        the track requirements;}
\item{Rejection of non-$ep$ background: the muon-rejection cuts described in 
      Section~\ref{ss:StanEvSel} were used. A class of background events arose
      from beam-halo muons that produced a shower inside the FCAL. To reduce
      this background, topological cuts on the transverse and longitudinal
      shower shape were imposed; these cuts rejected events in which the energy
      deposits were more collimated than for typical hadronic jets.}
\end{itemize}

\subsection{Final event sample}
\label{ss:FidBin}
To restrict the sample to kinematic regions where
the resolution in the kinematic variables was good and
the backgrounds small, the requirements $Q^2_{\rm{JB}}>200\gev^2$
and $y_{\rm{JB}}<0.9$ were imposed.

All events were visually inspected, and five cosmic-ray and halo-muon events were removed from the large-$\gamma_0$ sample and 11 from the small-$\gamma_0$ sample. A total of 1164 data events satisfied these criteria in the large-$\gamma_0$ sample and 292 in the small-$\gamma_0$ sample, in good agreement with the 1183 and 285 predicted by the MC simulation for the large- and small-$\gamma_0$ samples, respectively. 

\begin{figure}
  \begin{center}
    \includegraphics*[width=.9\textwidth]{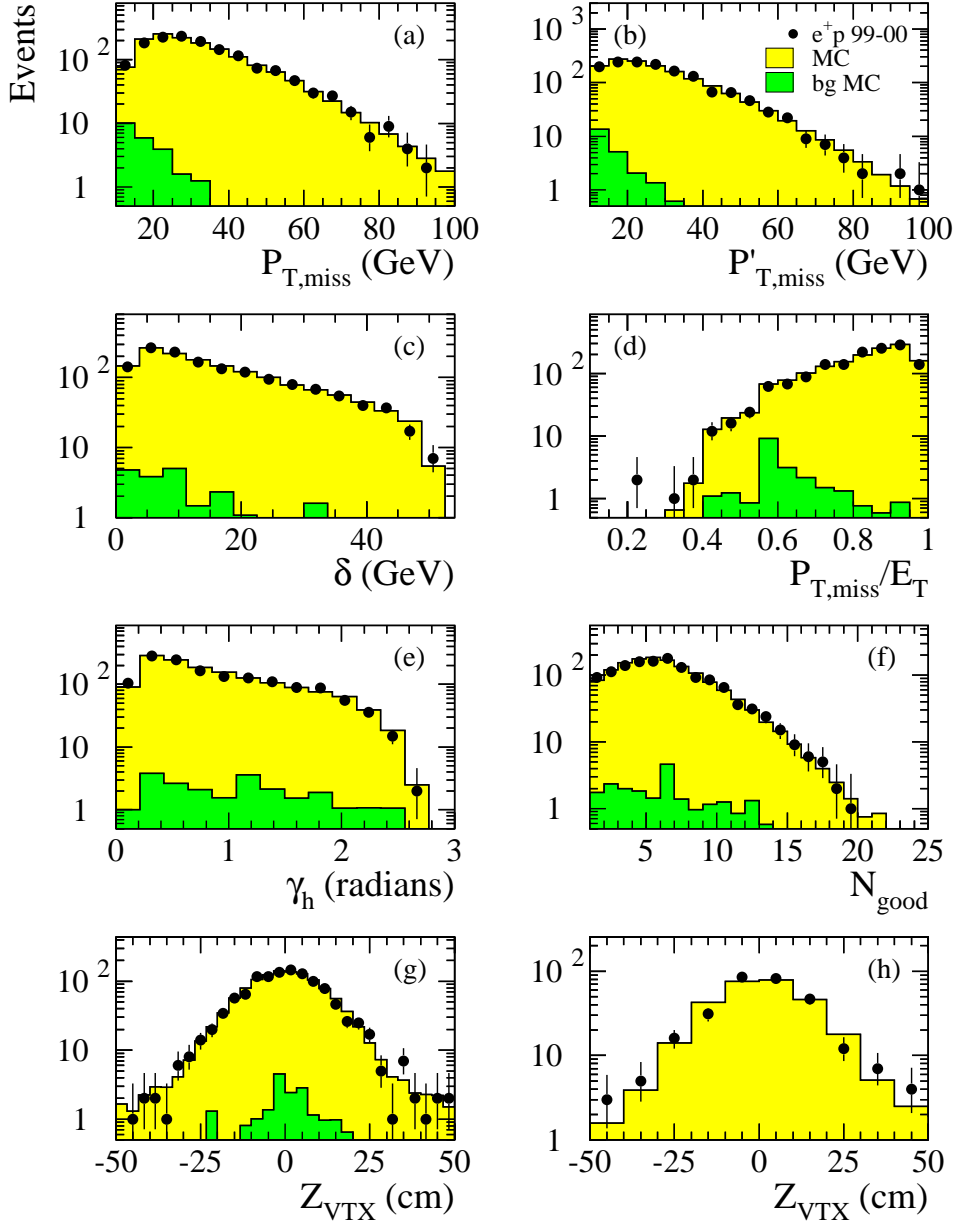}
 \end{center}
  \caption{
    Comparison of the final $e^+ p$ CC data sample (solid points) with the 
    sums of the signal and $ep$-background Monte Carlo 
    simulations (light shaded histograms). The simulated $ep$-background events
    are shown as the dark shaded histograms. Shown are the distributions of 
    (a) the missing transverse momentum,
    $\PTM$, (b) $\PTM$ excluding the very forward cells, $\PTM'$, 
    (c) the variable 
    $\delta$, defined in Section~\ref{s:reconstruction}, (d) 
    the ratio of missing transverse momentum to total transverse energy, 
    $\PTM/E_T$,
    (e) $\gamma_h$, (f) the number of good tracks,
    (g) the $Z$ position of the CTD vertex for the large-$\gamma_0$ sample 
and (h) the $Z$ position of the timing 
    vertex for the small-$\gamma_0$ sample. 
    }
  \label{f:control}
\end{figure}

Figure \ref{f:control} compares the distributions of data 
events entering the final CC sample with the MC expectation for the sum of the CC signal and $ep$ background events. 
The MC simulations give a good description of the data.
\section{Cross section determination and systematic \\ uncertainties}

\label{s:xsect}
The measured cross section
in a particular kinematic bin, for example for $d^2\sigma/dxdQ^2$, was determined from

\begin{equation}
  \frac{d^2\sigma}{dxdQ^2} = \frac{N_{\rm{data}}-N_{\rm{bg}}}{N_{\rm{MC}}} \cdot \frac{d^2 \sigma^{\rm{SM}}_{\rm{Born}}}{dxdQ^2}, \nonumber
\end{equation}

where $N_{\rm{data}}$ is the number of data events, $N_{\rm{bg}}$ is the number of background events estimated from the MC simulation and $N_{\rm{MC}}$ is the number of signal MC events. $\frac{d^2 \sigma^{\rm{SM}}_{\rm{Born}}}{dxdQ^2}$ is the Standard Model prediction evaluated in the on-shell scheme~\cite{epj:c15:1} using the PDG values for the electroweak parameters and the CTEQ5D PDFs~\cite{epj:c12:375}. A similar procedure was used for $d\sigma/dQ^2$, $d\sigma/dx$ and $d\sigma/dy$. 
Consequently, the acceptance, as well as the bin-centring and 
radiative corrections were all taken from the MC simulation. The cross sections $ d \sigma/ d  Q^2$ and $ d \sigma/ d  x$ were extrapolated to the full $y$ range using the SM predictions calculated with the CTEQ5D PDFs. The extrapolation factors ranged from 3\% at the lowest $Q^{2}$ to 28\% at the highest $Q^{2}$ in $ d \sigma/ d  Q^2$ and from zero to 8\% in $ d \sigma/ d  x$.

\label{s:systerr}
The systematic uncertainties in the measured cross sections were determined by
changing the selection cuts or the analysis procedure in turn and repeating the
extraction of the cross sections. 

\begin{itemize}
\item{Uncertainties of the calorimeter energy scale:  
\begin{itemize}
\item{the relative uncertainty of the hadronic energy scale was 
  2\% for the RCAL and 1\% for the FCAL and BCAL~\cite{pl:b539:197}. 
  Varying the energy scale of the calorimeter sections by these 
  amounts in the detector simulation induces  small  shifts of the Jacquet-Blondel 
  estimators of the kinematic variables. The variation of the energy scale of each 
  of the calorimeters simultaneously up or down
  by these amounts gave the systematic uncertainty 
  on the total measured energy in the calorimeter. This was found to give shifts 
  in the cross sections which were correlated between kinematic bins 
  ($\delta_{1}$ in Tables~\ref{t:single} and \ref{t:double});} 

\item{the uncertainty in the cross sections from the effect of the energy scale on the measurement 
  of $\gamma_{h}$ was obtained by increasing (decreasing) the FCAL and RCAL energy scales
  together while the BCAL energy scale was decreased (increased). This gave shifts 
  in the cross sections that were correlated between kinematic bins 
  ($\delta_{2}$ in Tables~\ref{t:single} and \ref{t:double});} 

\item{the sensitivity of the cross-section measurements to the fractions of the energy deposited in the 
  EMC and HAC sections of the calorimeter was determined by 
  simultaneously increasing the energy measured in the EMC section of the calorimeter by 2\% and
  decreasing the energy measured in the HAC section by 2\%, and vice-versa.
  This was done separately for each of the calorimeter sections;}                
\end{itemize}
  The final systematic error attributed to the uncertainty in the hadronic energy
  scale was obtained by taking the quadratic sum of these three types of estimate.
  The resulting systematic shifts in the measured cross sections were typically within 
  $\pm 5\%$, but increased to $\pm 20\%$ in the highest $Q^{2}$ and $x$ bins;}
\item{Energy leakage: of the accepted events, 4\% have a measurable energy leakage
  from the CAL into the BAC. The average leakage of transverse energy for these
  events is 5\% of that observed in the CAL. Both the fraction of events with
  leakage and the average amount of leakage are well modelled by the MC
  simulation and the effect on the cross-section measurement is negligible;}
\item{Variation of selection thresholds: the selection thresholds were varied
      by the typical resolution of the CAL quantities ($\pm10\%$) in both data
      and MC and no significant systematic effects were observed. The criteria
      for good tracks was also varied and no systematic effects in the measured 
      cross sections were observed;}
\item{Parton-shower scheme: the {\sc meps} model of {\sc lepto} 
    was used instead of the {\sc ariadne} model. This gave shifts in the cross
    sections which were found to be correlated between kinematic bins
    ($\delta_{3}$ in Tables~\ref{t:single} and
    \ref{t:double}). The largest effects were observed in the bins at
    high $Q^2$ ($\pm 20\%$) and  highest and lowest $x$ ($\pm 6\%$). The largest
    effect in the double-differential cross section was seen in the low-$Q^2$, 
    low-$x$ bins, where it amounted to $\pm 10\%$;}
\item{Background subtraction: the uncertainty in the small contribution from 
    photoproduction was estimated
    by fitting a linear combination of the $\PTM/E_T$ distributions
    of the signal and the background MC samples to the corresponding
    distribution in the data, allowing the normalisation ($N_{\rm{PhP}}$) of the 
    photoproduction MC events to vary. 
    No cut on $\PTM/E_T$ was applied for this check. 
    Varying $N_{\rm{PhP}}$ by $\pm 20\%$, corresponding to twice
    the uncertainty given by the fit, results in modifications of 
    the cross sections within $\pm 2\%$;}
\item{Parton distributions: the CC MC events were generated 
    using the CTEQ5D PDFs~\cite{epj:c12:375}.
    The ZEUS-S fit \cite{pr:d67:012007}
    was used to examine the influence of variations of the PDFs on the
    cross-section measurement through differences in the acceptance and
    bin-centring corrections. 
    The Monte Carlo events were reweighted to the extremes of the cross-section 
    prediction allowed by the fit.
    The change in the measured cross section was typically $<1\%$, 
    except at high $Q^{2}$ where it was $-5\%$ and at high $x$ where it 
    was $+4\%$;}
\item{NLO QCD corrections: the {\sc djangoh} program neglects the 
    $F_{\rm{L}}$ contribution and NLO QCD corrections to $xF_{3}$ 
    when generating CC events.
    The corresponding effect on the cross-section measurement was evaluated by
    reweighting the MC events with the ratio of the cross sections
    with and without NLO QCD corrections.  The largest effect, of $-3\%$, was 
    observed in the highest $y$ bin.}
\end{itemize}

The uncertainties associated with the trigger and the measurement of the vertex 
positions were negligible. The individual uncertainties were added in quadrature 
separately for the positive and negative deviations from the
nominal cross-section values to obtain the total systematic uncertainties 
listed in Tables \ref{t:single} and \ref{t:double}.
The $\mathcal{O}(\alpha)$ electroweak corrections to CC DIS 
have been discussed by several authors~\cite{jp:g25:1387,proc:mc:1998:530}. 
Various theoretical approximations and computer codes gave differences in 
the CC cross sections of typically $\pm(1-2)\%$ or less. 
 However, the differences can be as large as ~$\pm(3-8)\%$ at high $x$ and high $y$.  
This uncertainty and the uncertainty of $2.5\%$ on the measured total luminosity were
 not included in the figures and the tables of the cross sections.

\section{Results}
\label{ss:results}

The total cross section for $e^+ p$ CC DIS in the kinematic region $Q^2 >200\gev^2$ is
\begin{equation}
\sigma^{\rm{CC}}_{\rm{tot}} (Q^{2} > 200~\rm{\gev}^{2}) = 34.8 \pm 0.9(\rm{stat.}) ^{+0.9}_{-1.0}(\rm{syst.})~\rm{pb}. \nonumber
\end{equation}
In this case the uncertainty in the measured luminosity is included in the systematic uncertainty. The result is in agreement with the SM expectation of $37.0^{+1.7}_{-0.8}$~pb evaluated using the ZEUS-S fit.

\subsection{Single-differential cross sections}
\begin{figure}
  \begin{center}
    \includegraphics[width=.9\textwidth]{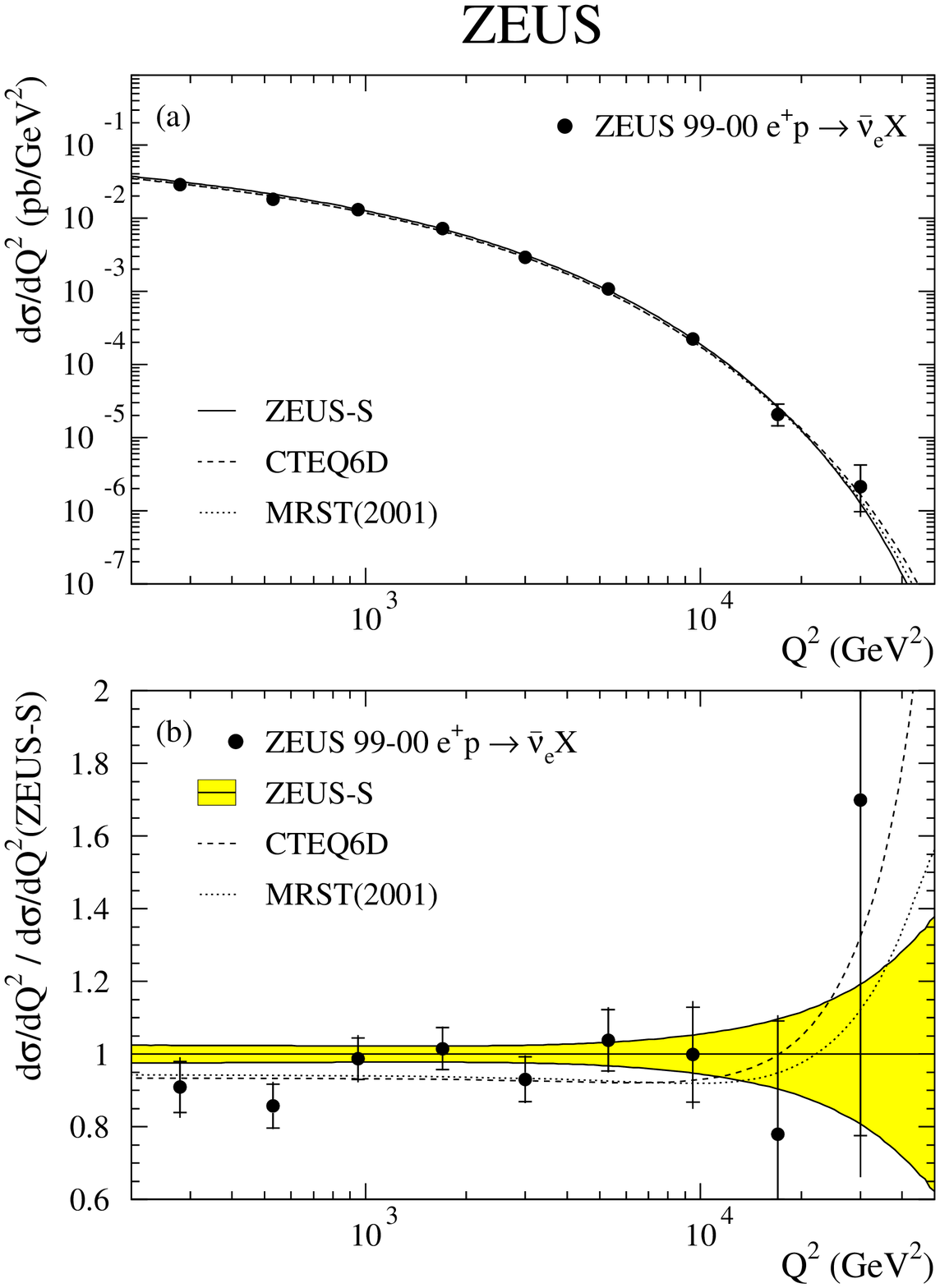}
  \end{center}
  \vskip -2.5mm
  \caption{
    (a) The $e^+p$ CC DIS Born cross section $d\sigma/dQ^2$ for data
    and the Standard Model expectation evaluated using
    the ZEUS-S, the CTEQ6D and the MRST (2001) PDFs.
    The data are shown as the filled points, the statistical uncertanties 
    are indicated by the inner error bars (delimited by horizontal lines) 
    and the full error bars show the total uncertainty obtained by adding 
    the statistical and systematic contributions in quadrature.
    (b) The ratio of the measured cross section, $d\sigma/dQ^2$, to the
    Standard Model expectation evaluated using the ZEUS-S fit.
    The shaded band shows the 
    uncertainties from the ZEUS-S fit.}
  \label{f:single_q2}
\end{figure}

\begin{figure}
  \begin{center}
    \includegraphics[width=.9\textwidth]{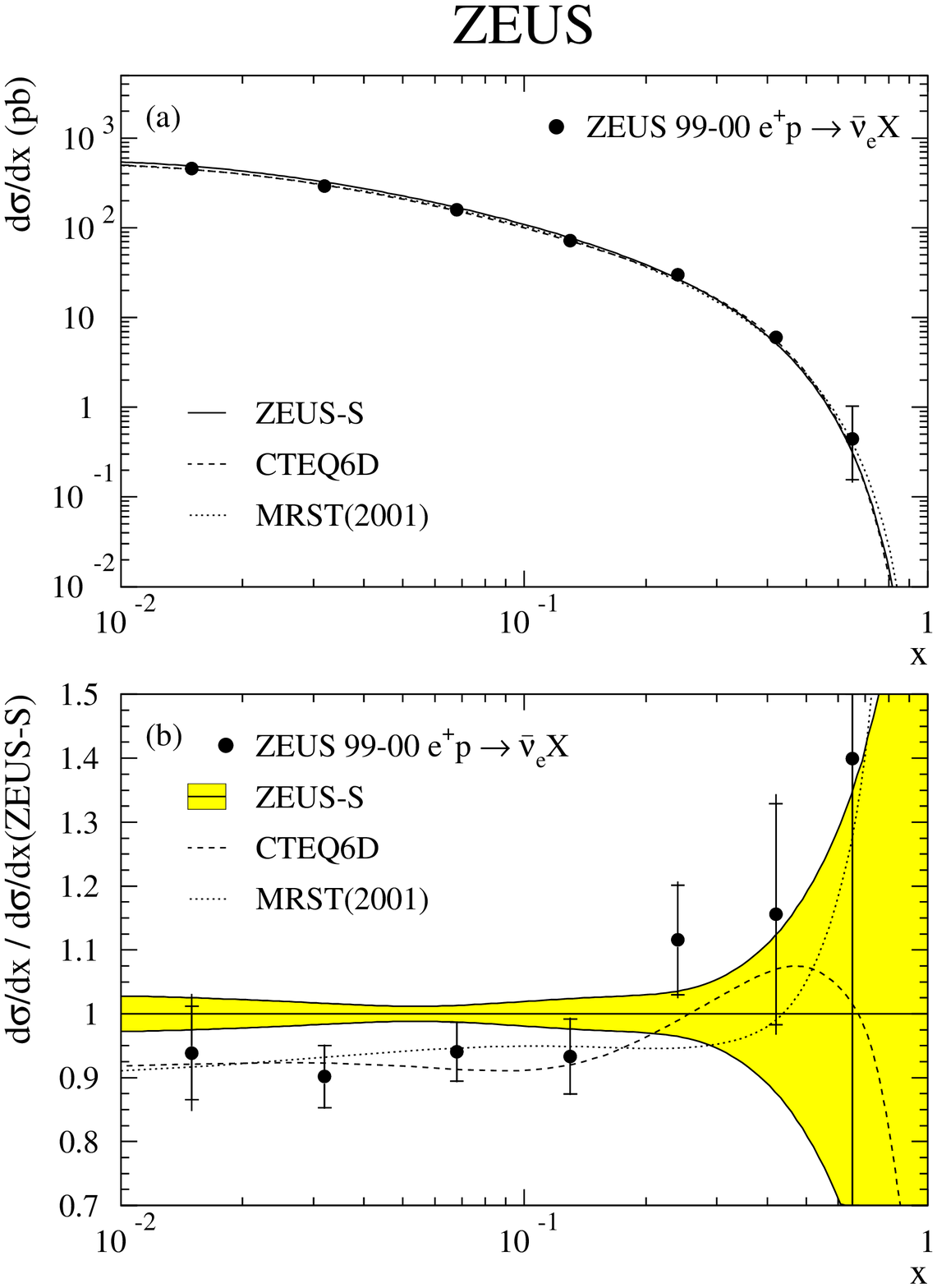}
  \end{center}
  \vskip -2.5mm
  \caption{
    (a) The $e^+p$ CC DIS Born cross section $d\sigma/dx$ for data
    and the Standard Model expectation evaluated using
    the ZEUS-S, the CTEQ6D and the MRST (2001) PDFs.
    The data are shown as the filled points, the statistical uncertanties 
    are indicated by the inner error bars (delimited by horizontal lines) 
    and the full error bars show the total uncertainty obtained by adding 
    the statistical and systematic contributions in quadrature.
    (b) The ratio of the measured cross section, $d\sigma/dx$, to the
    Standard Model expectation evaluated using the ZEUS-S fit.
    The shaded band shows the 
    uncertainties from the ZEUS-S fit.}
  \label{f:single_x}
\end{figure}

\begin{figure}
  \begin{center}
    \includegraphics[width=.9\textwidth]{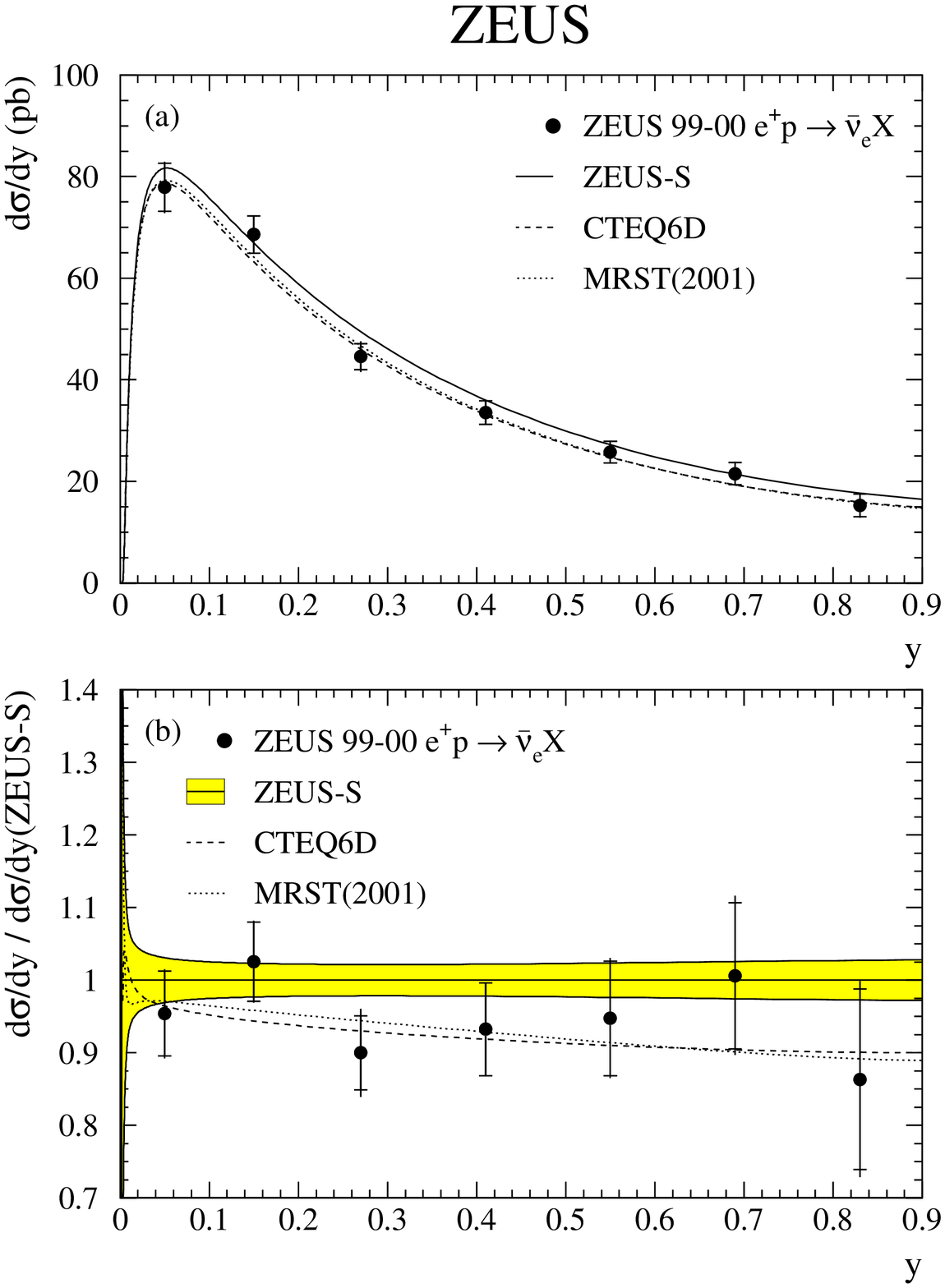}
  \end{center}
  \vskip -2.5mm
  \caption{
    (a) The $e^+p$ CC DIS Born cross section $d\sigma/dy$ for data
    and the Standard Model expectation evaluated using
    the ZEUS-S, the CTEQ6D and the MRST (2001) PDFs.
    The data are shown as the filled points, the statistical uncertanties 
    are indicated by the inner error bars (delimited by horizontal lines) 
    and the full error bars show the total uncertainty obtained by adding 
    the statistical and systematic contributions in quadrature.
    (b) The ratio of the measured cross section, $d\sigma/dy$, to the
    Standard Model expectation evaluated using the ZEUS-S fit.
    The shaded band shows the 
    uncertainties from the ZEUS-S fit.}
  \label{f:single_y}
\end{figure}

The single-differential cross sections $ d \sigma/ d  Q^2$, $ d \sigma/ d  x$ and $ d \sigma/ d  y$ for $Q^{2}>200~\gev^{2}$ are shown in Figs.~\ref{f:single_q2}, \ref{f:single_x} and \ref{f:single_y}, respectively, and compiled in Table~\ref{t:single} including details of the systematic uncertainties that are correlated between cross-section bins. 
The SM cross sections derived from Eq.~(\ref{e:Bornpos}) using the 
ZEUS-S fit, the CTEQ6D~\cite{jhep:07:012} and the  
MRST(2001)~\cite{epj:c23:73} 
parameterisations of the PDFs 
are shown, together with the ratios of the 
measured cross sections to the SM cross section evaluated with the 
ZEUS-S fit.

The cross sections $ d \sigma/ d  Q^2$ and $ d \sigma/ d  x$ drop by four and three
orders of magnitude, respectively, due to the effect of the $W$-boson propagator and the decreasing quark density at large $x$. 
The ZEUS-S fit was based on fixed-target DIS data obtained at much lower $Q^{2}$ ($<100\gev^{2}$) and from ZEUS NC data at large $Q^{2}$. 
The excellent description of the data by the SM prediction based on this fit
confirms both the decomposition of the proton momentum into different quark flavours, specifically the down-quark contribution, and the evolution of parton distributions towards scales considerably larger than the $W$-boson mass. 
At very large $x$ and $Q^{2}$, the uncertainty in the prediction derived from the ZEUS-S fit, and also the global fits, reflects the lack of data constraining the $d$-quark density.

\subsection{Double-differential cross sections}
\begin{figure}
  \begin{center}
    \includegraphics[width=0.9\textwidth]{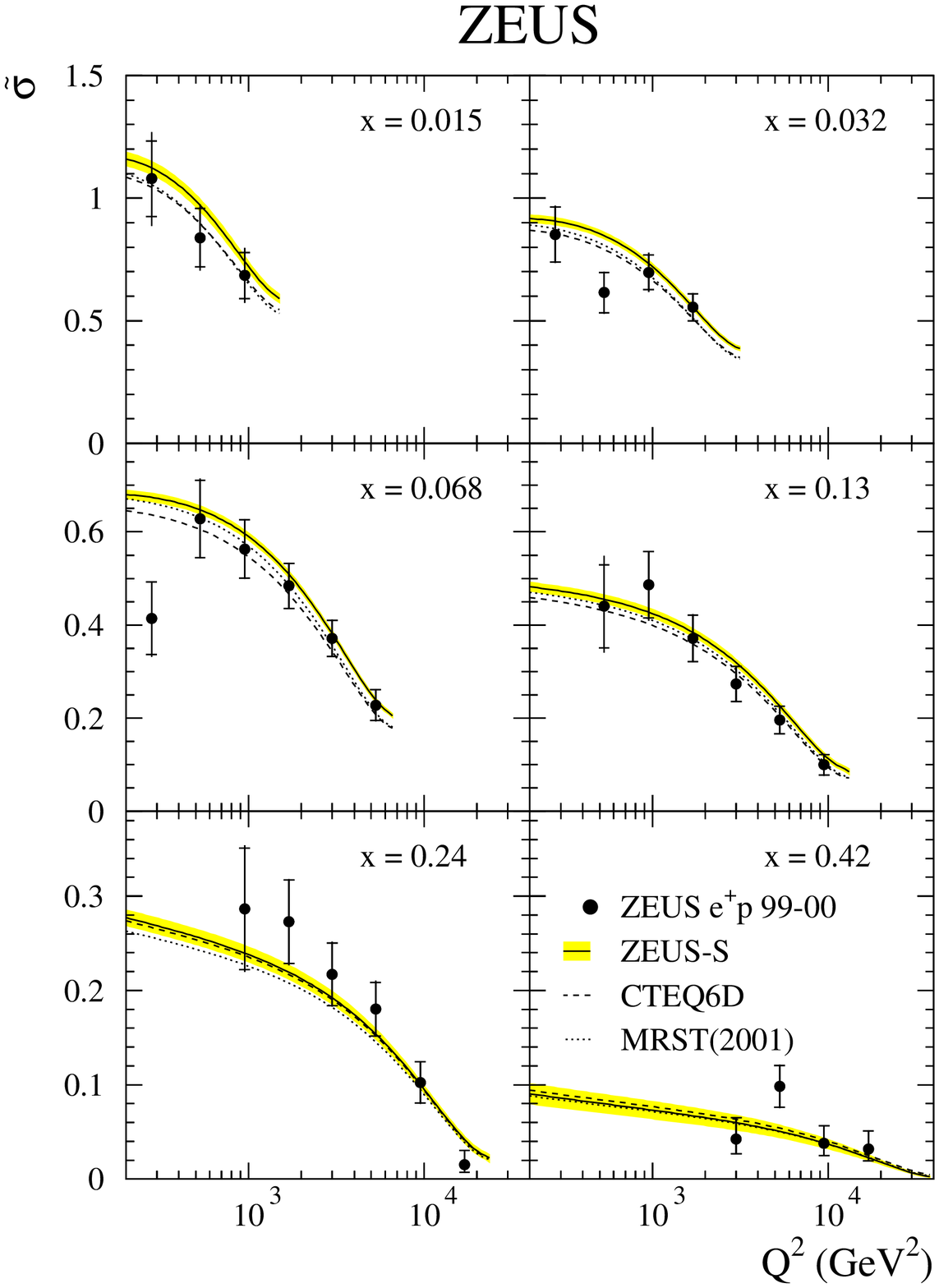}
  \end{center}
  \caption{
    The reduced cross section, $\tilde{\sigma}$, as a function of $Q^2$,
    for different fixed values of $x$.
    The data are shown as the filled points, the statistical uncertanties 
    are indicated by the inner error bars (delimited by horizontal lines) 
    and the full error bars show the total uncertainty obtained by adding 
    the statistical and systematic contributions in quadrature.
    The expectation of the
    Standard Model evaluated using 
    the ZEUS-S, the CTEQ6D and the MRST(2001) PDFs is shown by the solid, 
    dashed and dotted lines, respectively. The shaded band shows the uncertainties
    from the ZEUS-S fit.}
  \label{f:ddfixx}
\end{figure}

\begin{figure}
  \begin{center}
    \includegraphics[width=0.9\textwidth]{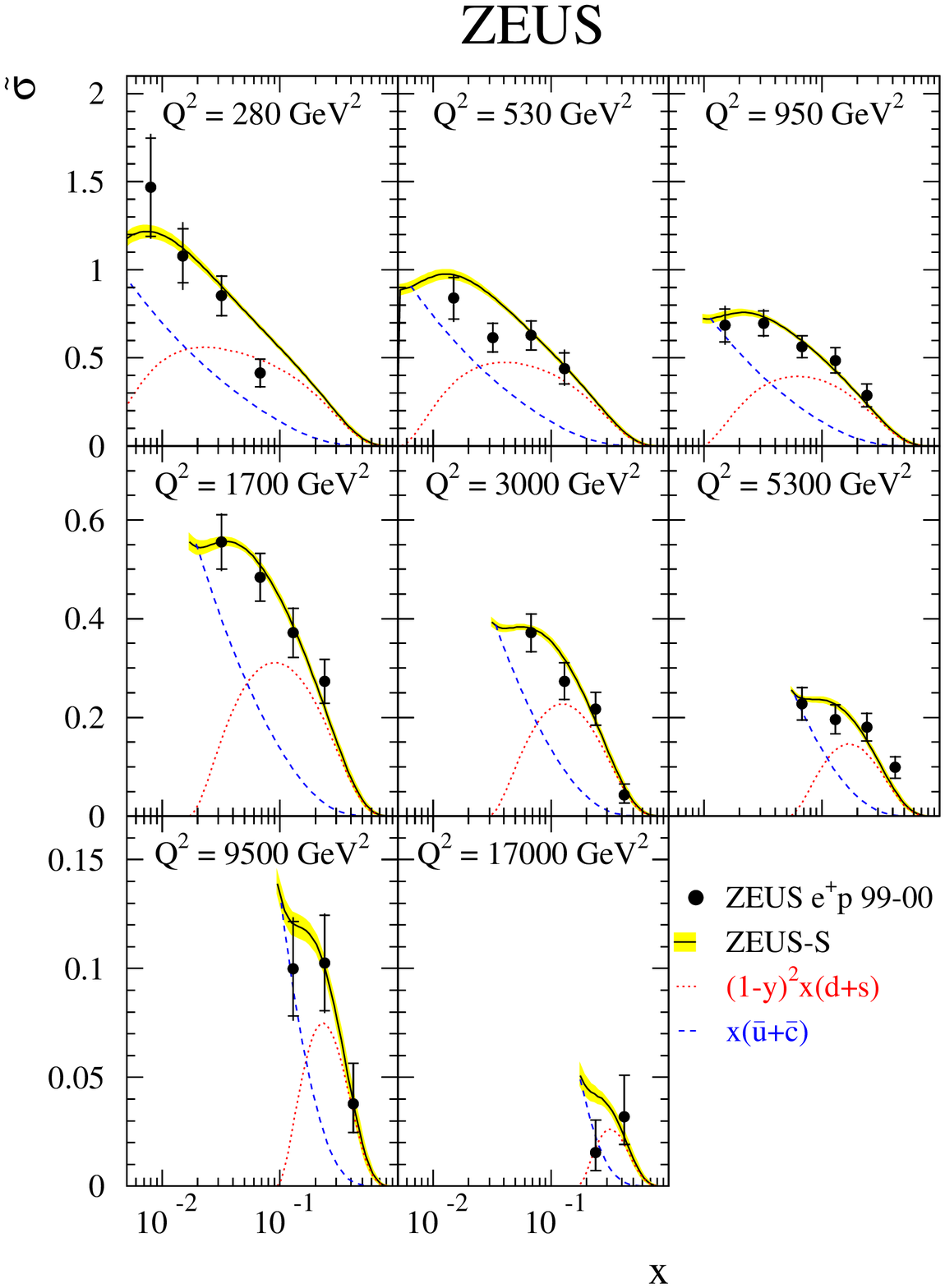}
  \end{center}
  \caption{
    The reduced cross section, $\tilde{\sigma}$, as a function of $x$,
    for different values of $Q^2$.
    The data are shown as the filled points, the statistical uncertanties 
    are indicated by the inner error bars (delimited by horizontal lines) 
    and the full error bars show the total uncertainty obtained by adding 
    the statistical and systematic contributions in quadrature.
    The expectation of the
    Standard Model evaluated using 
    the ZEUS-S fit
    is shown as a solid line. The shaded band shows the uncertainties
    from the ZEUS-S fit.
    The separate contributions of the PDF combinations $x (\bar{u}+\bar{c})$ and
    $(1-y)^2 x (d+s)$ are shown by the dotted  
    and dashed lines, respectively.
    }
  \label{f:ddfixq2}
\end{figure}

The reduced double-differential cross section, $\tilde{\sigma}$, is
defined by
\begin{equation}
   \tilde{\sigma} = \left[{G^2_F \over 2 \pi x } \Biggl( {M^2_W \over M^2_W + Q^2}\Biggr)^2 \right]^{-1}
{{d^2\sigma} \over {dx \, dQ^2}}. \nonumber
  \label{e:Reduced}
\end{equation}
At leading order in QCD, $\tilde{\sigma}({e^+ p \rightarrow \nubar_e X})$
depends on the quark momentum distributions as follows:
\begin{equation}
  \tilde{\sigma} (e^+ p \rightarrow {\nubar}_e X) = x\left[\bar{u} + \bar{c} + (1-y)^2 (d + s) \right].
  \label{e:LO}
\end{equation}
The reduced cross sections are displayed as functions of $Q^2$ and $x$ 
in Figs. \ref{f:ddfixx} and \ref{f:ddfixq2}, respectively, and compiled in Table~\ref{t:double} including details of the systematic uncertainties that are correlated between cross-section bins. 
The predictions of Eq.~(\ref{e:Bornpos}), evaluated using the
ZEUS-S fit, the CTEQ6D and the MRST(2001) PDFs give a good description of the data. 
The contributions from the PDF combinations $(d + s)$ and $(\bar{u} + \bar{c})$, obtained 
in the \MSbar scheme from the ZEUS-S fit, are shown separately in 
Fig.~\ref{f:ddfixq2}. 

\subsection{Helicity studies}
The $W$ boson couples only to left-handed fermions and right-handed antifermions. Therefore, the angular distribution of the scattered quark in $e^{+}\bar{q}$ CC DIS will be flat in $\theta^{*}$, while it will exhibit a $(1+\cos\theta^{*})^{2}$ distribution in $e^{+}q$ scattering. Since $(1-y)^2 \propto (1+\cos\theta^{*})^{2}$, the helicity structure of CC interactions can be illustrated 
by plotting the reduced double-differential cross section of Eq.~(\ref{e:LO}) 
versus $(1-y)^2$ in bins of $x$.
In the region of approximate scaling, i.e. $x\sim 0.1$,
this yields a straight line. At leading order in QCD, the intercept of this line gives 
the ($\bar{u}+\bar{c}$) contribution, while the slope gives the ($d+s$) 
contribution. 

\begin{figure}
  \begin{center}
    \includegraphics[width=.9\textwidth]{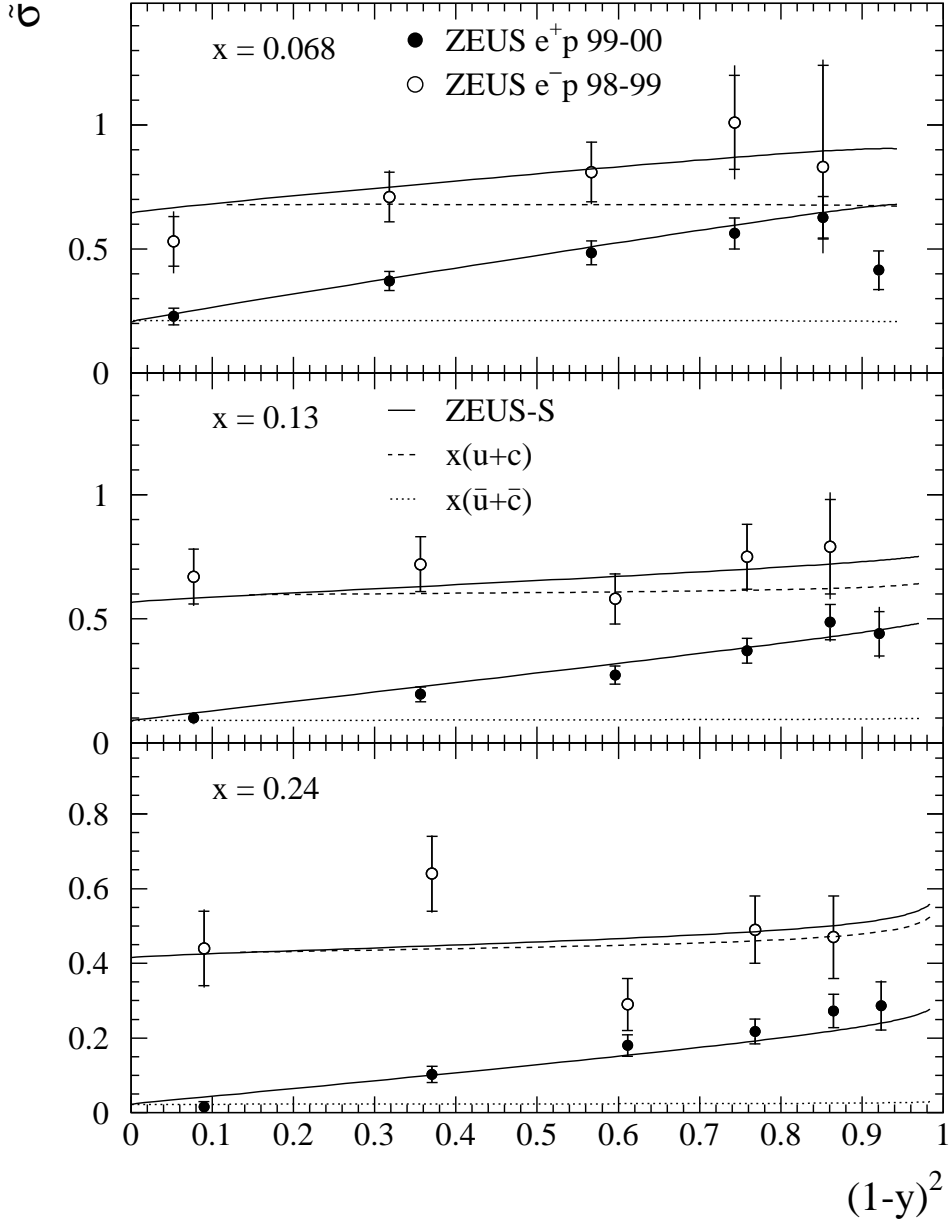}
  \end{center}
  \caption{
    The reduced cross section, $\tilde{\sigma}$, as a function of $(1-y)^2$,
    for different fixed values of $x$, for $e^+ p$ (solid points) and $e^- p$
    (open circles) CC DIS.
    The data are shown as the points, the statistical uncertanties 
    are indicated by the inner error bars (delimited by horizontal lines) 
    and the full error bars show the total uncertainty obtained by adding 
    the statistical and systematic contributions in quadrature.
    The expectation of the
    Standard Model evaluated using 
    the ZEUS-S fit is shown as a solid line.
    The contributions of the PDF combinations $x (u+c)$ and
    $x (\bar{u}+\bar{c})$ are shown by the dashed 
    and dotted lines, respectively.}
  \label{f:helicity}
\end{figure}

Figure \ref{f:helicity} shows $\tilde{\sigma}$ as a function of $(1-y)^2$ for the $e^+ p$ CC DIS data, 
compared to the previously published $e^- p$ data~\cite{pl:b539:197}. 
At large $x$, $e^+ p$ CC DIS is sensitive to the valence part of $d(x,Q^2 )$, while $e^- p$ CC DIS is 
sensitive to the valence part of $u(x,Q^2 )$. 
Equivalently to Eq.~(\ref{e:LO}), the reduced cross section 
for $e^- p$ CC DIS can be written as 
\begin{equation}
  \tilde{\sigma}(e^- p \rightarrow \nu_e X) = x\left[u + c + (1-y)^2 (\bar{d} + \bar{s}) \right], \nonumber
\end{equation}
permitting a similar interpretation for the intercept and slope in terms of
the appropriate parton densities. Scaling violations can be observed in the theoretical prediction as $(1-y)^2$ approaches one.
The data agree with the expectation of the SM from the ZEUS-S fit. 

\subsection{Mass of the \boldmath{$W$} boson}
\label{ss:MW}
The fall in the cross section $ d \sigma/ d  Q^2$ with increasing $Q^2$ depends on $M_{W}^{4}/(Q^{2}+M_{W}^{2})^{2}$. 
Fitting $ d \sigma/ d  Q^2$ with $G_F$ fixed at the PDG\cite{epj:c15:1} 
value of $1.16639 \cdot 10^{-5}\gev^{-2}$, using the ZEUS-S fit PDFs and $M_W$ treated as a free parameter, gives:

\begin{equation}
M_W=78.9\pm 2.0{\rm (stat.)}\pm 1.8{\rm(syst.)} ^{+2.0}_{-1.8}\,{\rm(PDF)}\,\gev,\nonumber
\end{equation}

where the third uncertainty was estimated by varying the PDFs within 
the uncertainties given by the ZEUS-S fit. 
The systematic uncertainty includes contributions from the sources identified 
in Section \ref{s:systerr} and the uncertainty on the measured 
luminosity. 
This measurement, in the space-like region, is in good agreement with 
the more precise measurements of $W$-boson mass in the time-like region~\cite{epj:c15:1}.

\subsection{Extraction of \boldmath$F_{2}^{\rm{CC}}$}
\label{ss:f2}
\begin{figure}
  \begin{center}
    \includegraphics[width=.9\textwidth]{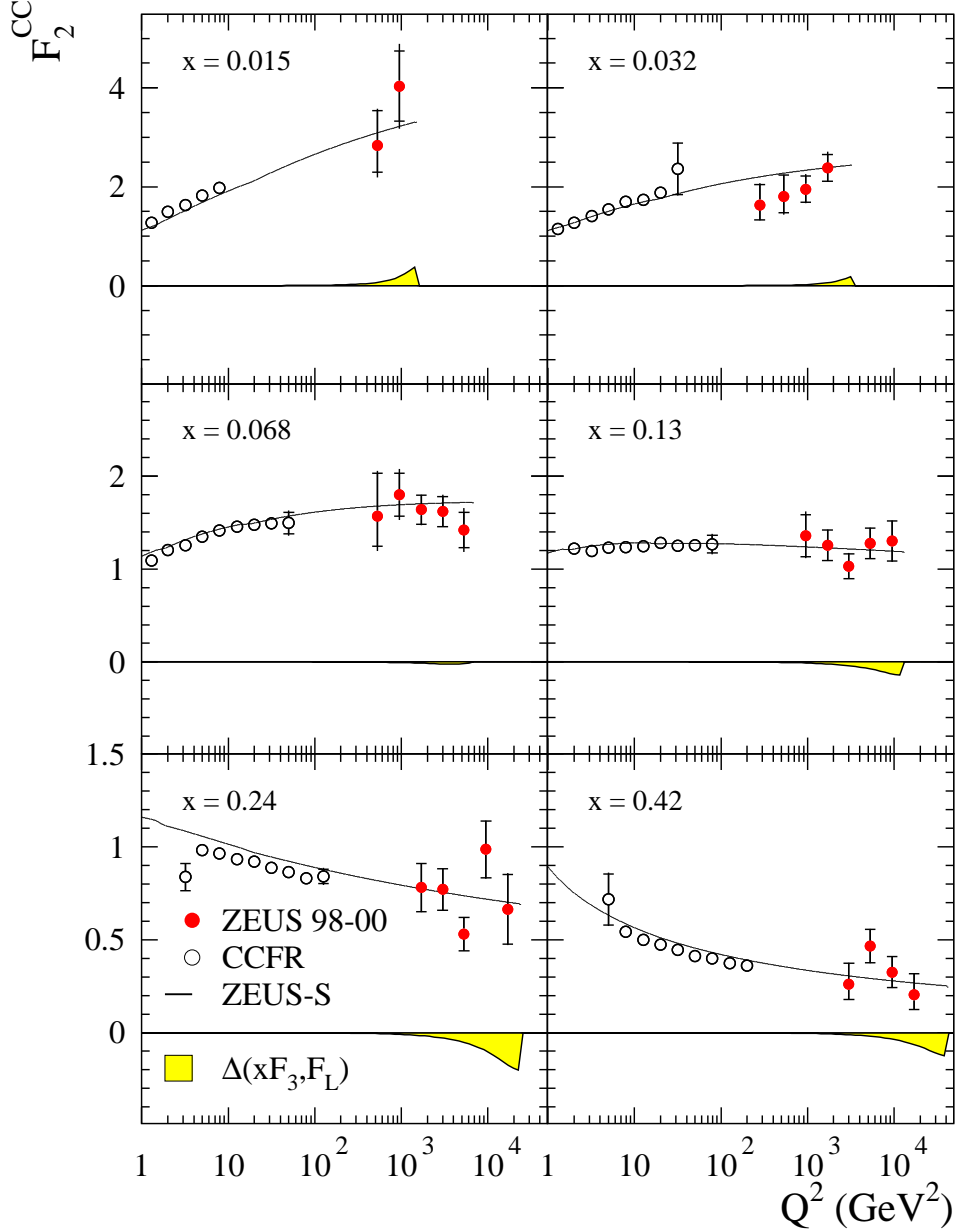}
  \end{center}
  \caption{
    The structure-function $F_{2}^{\rm{CC}}$ as a function of $Q^{2}$,
    for different fixed values of $x$, extracted from ZEUS data (solid points) 
    and compared to CCFR measurements corrected for heavy-target effects (open circles).
    The expectation of the
    Standard Model evaluated using 
    the ZEUS-S fit is shown as a solid line. The shaded band shows the size of the correction
    $\Delta (xF_{3},F_{\rm{L}})$ defined in Section~\ref{ss:f2}.}
  \label{f:f2}
\end{figure}

It is possible to extract the quark singlet distribution, $F_{2}^{\rm{CC}}$, by combining the measurements presented in this paper with the previous ZEUS $e^- p$ CC DIS results~\cite{pl:b539:197}. This can be compared to the corresponding result from fixed-target neutrino-nuleon scattering.
Equations \ref{e:f2pos} and \ref{e:xf3pos} show that only down-type quarks and up-type antiquarks contribute to $e^{+}p$ CC DIS in the Standard Model. The cross section for $e^{-}p$ CC DIS is given by
\begin{equation}
\frac{d^{2}\sigma^{\rm{CC}}_{Born}(e^{-}p)}{dxdQ^{2}} = \frac{G^{2}_{\rm{F}}}{4\pi x}
\frac{M_{W}^{4}}{(Q^{2}+M_{W}^{2})^{2}}[Y_{+}F_{2,e^{-}p}^{\rm{CC}}(x,Q^{2})-y^{2}F_{\rm{L},e^{-}p}^
{\rm{CC}}(x,Q^{2}) + Y_{-}xF_{3,e^{-}p}^{\rm{CC}}(x,Q^{2})], \nonumber
\label{e:Bornele}
\end{equation}
where the structure-functions $F_{2,e^{-}p}^{\rm{CC}}$ and $xF_{3,e^{-}p}^{\rm{CC}}$, at leading order in QCD, may be written in terms of sums and differences of quark and antiquark PDFs:
\begin{equation}
F_{2,e^{-}p}^{\rm{CC}} = x[u(x,Q^{2})+c(x,Q^{2}) +\bar{d}(x,Q^{2})+\bar{s}(x,Q^{2})],  \nonumber
\label{e:f2ele}
\end{equation}
\begin{equation}
xF_{3,e^{-}p}^{\rm{CC}} = x[u(x,Q^{2})+c(x,Q^{2})-\bar{d}(x,Q^{2})-\bar{s}(x,Q^{2})]. \nonumber
\label{e:xf3ele}
\end{equation}
Therefore, the sum of $F_{2,e^{+}p}^{\rm{CC}}$ and $F_{2,e^{-}p}^{\rm{CC}}$ represents the contribution from all quark and antiquark flavours and can be extracted from the measured CC cross sections as 
\begin{equation}
F_{2}^{\rm{CC}}= \frac{4\pi x}{G_{F}^{2}}\bigg( \frac{M_{W}^{2}+Q^{2}}{M_{W}^{2}} \bigg)^{2} \frac{1}{Y_{+}} \bigg( \frac{d^{2}\sigma^{\rm{CC}}_{Born}(e^{+}p)}{dxdQ^{2}} + \frac{d^{2}\sigma^{\rm{CC}}_{Born}(e^{-}p)}{dxdQ^{2}}\bigg) + \Delta(xF_{3},F_{\rm{L}}), \nonumber
\end{equation}
where $\Delta(xF_{3},F_{\rm{L}})$ denotes a correction term taking into account the $xF_{3}$ and $F_{\rm{L}}$ structure functions. The correction is given by
\begin{equation}
\Delta(xF_{3},F_{\rm{L}}) = \frac{Y_{-}}{Y_{+}}\bigg( xF_{3,e^{+}p}^{\rm{CC}} - xF_{3,e^{-}p}^{\rm{CC}} \bigg)+\frac{y^{2}}{Y_{+}}\bigg(F_{{\rm{L}},e^{+}p}^{\rm{CC}}+F_{{\rm{L}},e^{-}p}^{\rm{CC}} \bigg).  \nonumber
\end{equation}
The dominant uncertainty on the extracted structure-function is statistical. The
systematic uncertainties $\delta_{1}$, $\delta_{2}$, $\delta_{3}$ (defined in
Section~\ref{s:systerr}) and the luminosity uncertainties were considered fully correlated between the two data sets. The other systematic uncertainties were treated as uncorrelated. Figure \ref{f:f2} shows the extracted structure-function $F_{2}^{\rm{CC}}$ as a function of $Q^{2}$ for different values of $x$. The size of the correction $\Delta(xF_{3},F_{\rm{L}})$, computed at NLO in QCD using the ZEUS-S fit, is shown as a shaded area in each bin. It can be seen that the correction term is smaller than the uncertainties on the measurement for low values of $x$ but becomes sizeable at higher values of $x$ and $Q^{2}$. The uncertainity on the correction $\Delta(xF_{3},F_{\rm{L}})$ was also computed using the ZEUS-S fit and found to be small. The corresponding result for $\nu \rm{Fe}$ interactions from the CCFR Collaboration~\cite{prl:86:2742} is shown, after correcting for heavy-target effects~\cite{npps:b79:105}. It can be seen that both sets of results, spanning more than four orders of magnitude in $Q^2$, are well described by the SM prediction evaluated using the ZEUS-S fit. The prediction is somewhat above the CCFR data in the highest x bins. Note that the CCFR $F_{2}^{\rm{CC}}$ results were not included in the ZEUS-S fit. The global fits MRST(2001) and CTEQ6D which use the CCFR results obtain a reasonable description of both data sets over the entire kinematic range.

\section{Summary}
\label{s:summary}
Differential cross sections for charged current deep inelastic scattering,
\mbox{$e^+p \rightarrow \nubar_{e} X$}, have been measured for $Q^{2}>200~\gev^{2}$ 
using $60.9~\rm{pb^{-1}}$ of data  collected with the ZEUS detector during the 
period 1999 to 2000. 
The double-differential cross section $d^2\sigma/dx\,dQ^2$ is presented 
in the kinematic range \mbox{$280\gev^2 <Q^2<17~000 \gev^2$} and
\mbox{$0.008<x<0.42$}. The chiral structure of the Standard Model was
investigated by plotting the double-differential cross section as a function of
$(1-y)^2$. The mass of the $W$ boson, determined from a fit to $d \sigma/ d
Q^2$, is \mbox{$M_W=78.9\pm 2.0\,{\rm (stat.)}\,\pm 1.8\,{\rm(syst.)}\,
^{+2.0}_{-1.8}\,{\rm(PDF)}\,\gev$}. The singlet structure-function
$F_{2}^{\rm{CC}}$ has been extracted for the first time at HERA, by combining
the measurements presented here with previous ZEUS measurements. The Standard
Model gives an excellent description of all data, confirming the
decomposition of the proton momentum into different quark flavours, specifically
the down-quark distribution, and also verifies the QCD evolution of parton 
distributions towards scales considerably larger than the $W$-boson mass. 

\section*{Acknowledgements}
We appreciate the contributions to the construction and maintenance
of the ZEUS detector of the many people who are not listed as authors.
The HERA machine group and the DESY computing staff are especially
acknowledged for their success in providing excellent operation of the
collider and the data-analysis environment.
We thank the DESY directorate for their strong support and encouragement.

\vfill\eject

\def\bibname{\Large\bf References}
\def\refname{\Large\bf References}
\pagestyle{plain}
\ifzeusbst
  \bibliographystyle{./BiBTeX/bst/l4z_default}
\fi
\ifzdrftbst
  \bibliographystyle{./BiBTeX/bst/l4z_draft}
\fi
\ifzbstepj
  \bibliographystyle{./BiBTeX/bst/l4z_epj}
\fi
\ifzbstnp
  \bibliographystyle{./BiBTeX/bst/l4z_np}
\fi
\ifzbstpl
  \bibliographystyle{./BiBTeX/bst/l4z_pl}
\fi
{\raggedright
\bibliography{./BiBTeX/user/syn.bib,%
              ./BiBTeX/bib/l4z_articles.bib,%
              ./BiBTeX/bib/l4z_books.bib,%
              ./BiBTeX/bib/l4z_conferences.bib,%
              ./BiBTeX/bib/l4z_h1.bib,%
              ./BiBTeX/bib/l4z_misc.bib,%
              ./BiBTeX/bib/l4z_old.bib,%
              ./BiBTeX/bib/l4z_preprints.bib,%
              ./BiBTeX/bib/l4z_replaced.bib,%
              ./BiBTeX/bib/l4z_temporary.bib,%
              ./BiBTeX/bib/l4z_zeus.bib}}
\vfill\eject


\begin{table}
\scriptsize
\begin{center}
    \begin{tabular}{|c|c|c|c|c|c|c|c|c|}
    \hline
\multicolumn{9}{|c|}{$d\sigma/dQ^2$} \\
    \hline
$Q^2$ range & 
$Q^2$ & 
$N_{\scriptsize{\textrm{data}}}$ &
$N_{\scriptsize{\textrm{bg}}}$ & 
$\delta_{\scriptsize{\textrm{unc}}}(\%)$ & 
$\delta_{1}(\%)$ & 
$\delta_{2}(\%)$ & 
$\delta_{3}(\%)$ & 
$d\sigma/dQ^2$ (pb/GeV$^2$) \\ 

(\gev$^2$) & 
(\gev$^2$) & 
&
& 
& 
& 
& 
& 
\\ 
    \hline
     200~-     400 &      280 & 
  159 & 14.2 & 
$ ~^{+   2.0}_{-  3.1} $& 
$ ~^{+   1.9}_{-  1.8} $& 
$ ~^{+   0.1}_{-  0.5} $& 
$ ~^{-   3.8}_{+  3.8} $& 
$(2.85\pm0.22^{+0.13}_{-0.15}) \cdot 10^{ -2}$ \\ 
     400~-     711 &      530 & 
  204 &  2.3 & 
$ ~^{+   1.8}_{-  1.5} $& 
$ ~^{+   1.7}_{-  0.8} $& 
$ ~^{+   0.8}_{+  0.8} $& 
$ ~^{-   1.6}_{+  1.6} $& 
$(1.81\pm0.13^{+0.06}_{-0.04}) \cdot 10^{ -2}$ \\ 
     711~-    1265 &      950 & 
  306 &  4.2 & 
$ ~^{+   1.0}_{-  1.2} $& 
$ ~^{+   0.7}_{-  1.0} $& 
$ ~^{-   0.5}_{-  0.5} $& 
$ ~^{-   2.7}_{+  2.7} $& 
$(1.30\pm0.08^{+0.04}_{-0.04}) \cdot 10^{ -2}$ \\ 
    1265~-    2249 &     1700 & 
  324 &  1.4 & 
$ ~^{+   1.2}_{-  0.9} $& 
$ ~^{+   0.5}_{-  0.0} $& 
$ ~^{+   0.3}_{+  0.2} $& 
$ ~^{+   1.6}_{-  1.6} $& 
$(7.16\pm0.41^{+0.15}_{-0.13}) \cdot 10^{ -3}$ \\ 
    2249~-    4000 &     3000 & 
  235 &  0.9 & 
$ ~^{+   1.1}_{-  1.0} $& 
$ ~^{-   1.9}_{+  1.8} $& 
$ ~^{-   0.1}_{+  0.2} $& 
$ ~^{-   0.6}_{+  0.6} $& 
$(2.90\pm0.19^{+0.06}_{-0.07}) \cdot 10^{ -3}$ \\ 
    4000~-    7113 &     5300 & 
  155 &  0.2 & 
$ ~^{+   0.8}_{-  1.1} $& 
$ ~^{-   2.2}_{+  2.9} $& 
$ ~^{+   0.1}_{+  0.6} $& 
$ ~^{+   0.4}_{-  0.4} $& 
$(1.07\pm0.09^{+0.03}_{-0.03}) \cdot 10^{ -3}$ \\ 
    7113~-   12649 &     9500 & 
   59 &  0.2 & 
$ ~^{+   1.8}_{-  1.1} $& 
$ ~^{-   6.4}_{+  5.9} $& 
$ ~^{+   0.3}_{-  1.9} $& 
$ ~^{+   2.6}_{-  2.6} $& 
$(2.20\pm0.29^{+0.15}_{-0.16}) \cdot 10^{ -4}$ \\ 
   12649~-   22494 &    17000 & 
   11 &  0.0 & 
$ ~^{+   1.5}_{-  2.7} $& 
$ ~^{-   8.8}_{+   11} $& 
$ ~^{+   2.5}_{-  2.4} $& 
$ ~^{+   6.5}_{-  6.5} $& 
$(2.05^{+0.82}_{-0.61}~^{+0.26}_{-0.23}) \cdot 10^{ -5}$ \\ 
   22494~-   60000 &    30000 & 
    3 &  0.0 & 
$ ~^{+   4.7}_{-  6.3} $& 
$ ~^{-    16}_{+   18} $& 
$ ~^{+   5.1}_{-  7.8} $& 
$ ~^{+    21}_{-   21} $& 
$(2.12^{+2.06}_{-1.15}~^{+0.60}_{-0.59}) \cdot 10^{ -6}$ \\ 
    \hline
\multicolumn{9}{|c|}{$d\sigma/dx$} \\
    \hline
$x$ range & 
$x$ & 
$N_{\scriptsize{\textrm{data}}}$ & 
$N_{\scriptsize{\textrm{bg}}}$ & 
$\delta_{\scriptsize{\textrm{unc}}}(\%)$ & 
$\delta_{1}(\%)$ & 
$\delta_{2}(\%)$ & 
$\delta_{3}(\%)$ & 
$d\sigma/dx$ (pb) \\
    \hline
0.010~-0.022 & 0.015 & 
  167 &  7.7 & 
$ ~^{+   1.0}_{-  1.3} $& 
$ ~^{+   2.2}_{-  0.9} $& 
$ ~^{+   0.1}_{-  0.0} $& 
$ ~^{-   5.5}_{+  5.5} $& 
$(4.58\pm0.36^{+0.28}_{-0.26}) \cdot 10^{  2}$ \\ 
0.022~-0.046 & 0.032 & 
  351 &  6.2 & 
$ ~^{+   0.5}_{-  0.7} $& 
$ ~^{+   0.4}_{-  0.4} $& 
$ ~^{+   0.1}_{+  0.2} $& 
$ ~^{+   0.9}_{-  0.9} $& 
$(2.92\pm0.16^{+0.03}_{-0.04}) \cdot 10^{  2}$ \\ 
0.046~-0.100 & 0.068 & 
  425 &  4.8 & 
$ ~^{+   0.6}_{-  0.9} $& 
$ ~^{-   0.5}_{+  0.6} $& 
$ ~^{+   0.7}_{-  0.6} $& 
$ ~^{+   0.3}_{-  0.3} $& 
$(1.59\pm0.08^{+0.02}_{-0.02}) \cdot 10^{  2}$ \\ 
0.100~-0.178 & 0.130 & 
  258 &  0.9 & 
$ ~^{+   0.6}_{-  0.6} $& 
$ ~^{-   0.8}_{+  1.0} $& 
$ ~^{+   0.9}_{-  0.6} $& 
$ ~^{+   0.9}_{-  0.9} $& 
$(7.22\pm0.45^{+0.13}_{-0.11}) \cdot 10^{  1}$ \\ 
0.178~-0.316 & 0.240 & 
  173 &  0.5 & 
$ ~^{+   0.8}_{-  0.5} $& 
$ ~^{-   2.2}_{+  2.4} $& 
$ ~^{-   1.3}_{+  1.3} $& 
$ ~^{+   0.7}_{-  0.7} $& 
$(3.01\pm0.23^{+0.09}_{-0.08}) \cdot 10^{  1}$ \\ 
0.316~-0.562 & 0.420 & 
   45 &  0.1 & 
$ ~^{+   1.1}_{-  0.8} $& 
$ ~^{-   4.3}_{+  4.1} $& 
$ ~^{-   3.2}_{+  3.1} $& 
$ ~^{-   3.7}_{+  3.7} $& 
$(5.98\pm0.90^{+0.38}_{-0.39}) \cdot 10^{  0}$ \\ 
0.562~-1.000 & 0.650 & 
    2 &  0.0 & 
$ ~^{+   6.5}_{-  2.1} $& 
$ ~^{-    12}_{+   14} $& 
$ ~^{-    11}_{+   16} $& 
$ ~^{-   8.7}_{+  8.7} $& 
$(4.43^{+5.84}_{-2.87}~^{+1.06}_{-0.81}) \cdot 10^{ -1}$ \\ 
    \hline
\multicolumn{9}{|c|}{$d\sigma/dy$} \\
    \hline
$y$ range & 
$y$ & 
$N_{\scriptsize{\textrm{data}}}$ &
$N_{\scriptsize{\textrm{bg}}}$ & 
$\delta_{\scriptsize{\textrm{unc}}}(\%)$ & 
$\delta_{1}(\%)$ & 
$\delta_{2}(\%)$ & 
$\delta_{3}(\%)$ & 
$d\sigma/dy$ (pb) \\ 
    \hline
 0.00~- 0.10 &  0.05 & 
  264 &  7.0 & 
$ ~^{+   0.8}_{-  1.2} $& 
$ ~^{+   0.5}_{-  0.6} $& 
$ ~^{+   0.2}_{-  0.1} $& 
$ ~^{+   1.7}_{-  1.7} $& 
$(7.79\pm0.48^{+0.15}_{-0.17}) \cdot 10^{  1}$ \\ 
 0.10~- 0.20 &  0.15 & 
  360 &  6.3 & 
$ ~^{+   0.5}_{-  0.8} $& 
$ ~^{+   0.4}_{-  0.2} $& 
$ ~^{-   0.3}_{+  0.4} $& 
$ ~^{-   1.0}_{+  1.0} $& 
$(6.86\pm0.37^{+0.08}_{-0.09}) \cdot 10^{  1}$ \\ 
 0.20~- 0.34 &  0.27 & 
  316 &  4.1 & 
$ ~^{+   0.5}_{-  0.8} $& 
$ ~^{-   0.3}_{+  0.2} $& 
$ ~^{-   0.8}_{+  0.7} $& 
$ ~^{-   3.5}_{+  3.5} $& 
$(4.46\pm0.25^{+0.16}_{-0.16}) \cdot 10^{  1}$ \\ 
 0.34~- 0.48 &  0.41 & 
  219 &  1.9 & 
$ ~^{+   0.7}_{-  0.6} $& 
$ ~^{-   0.0}_{+  0.5} $& 
$ ~^{-   0.1}_{+  0.5} $& 
$ ~^{-   0.2}_{+  0.2} $& 
$(3.36\pm0.23^{+0.03}_{-0.02}) \cdot 10^{  1}$ \\ 
 0.48~- 0.62 &  0.55 & 
  146 &  2.3 & 
$ ~^{+   1.0}_{-  1.0} $& 
$ ~^{-   0.2}_{+  0.7} $& 
$ ~^{+   1.0}_{-  0.5} $& 
$ ~^{-   2.2}_{+  2.2} $& 
$(2.58\pm0.21^{+0.07}_{-0.06}) \cdot 10^{  1}$ \\ 
 0.62~- 0.76 &  0.69 & 
  102 &  1.4 & 
$ ~^{+   1.5}_{-  1.5} $& 
$ ~^{-   2.3}_{+  3.1} $& 
$ ~^{+   2.6}_{-  2.5} $& 
$ ~^{-   1.8}_{+  1.8} $& 
$(2.15\pm0.22^{+0.10}_{-0.09}) \cdot 10^{  1}$ \\ 
 0.76~- 0.90 &  0.83 & 
   49 &  0.4 & 
$ ~^{+   3.0}_{-  3.0} $& 
$ ~^{-   4.2}_{+  4.3} $& 
$ ~^{+   2.9}_{-  2.4} $& 
$ ~^{-   1.8}_{+  1.8} $& 
$(1.53\pm0.22^{+0.10}_{-0.09}) \cdot 10^{  1}$ \\ 
    \hline
    \end{tabular}
\end{center}
\caption{Values of the differential cross sections $d \sigma /dQ^{2}$, $d \sigma /dx$ and $d \sigma /dy$. The following quantities are given: the kinematic range of the measurement; the value of $x$,$y$ and $Q^2$ at which the cross section is quoted; the number of data events; the number of expected background events; the uncorrelated systematic uncertainty; those systematic uncertainties with correlations between cross-section bins $\delta_1$, $\delta_2$ and $\delta_3$ defined in Section~\ref{s:systerr} and the measured cross section, with statistical and total systematic uncertainties. The uncertainty on the measured luminosity of $2.5\%$ is not included in the total systematic uncertainty.}
  \label{t:single}
\end{table}

\begin{table}
\scriptsize
\begin{center}
    \begin{tabular}{|c|c|c|c|c|c|c|c|c|}
    \hline
$Q^2$ ( GeV$^2$ ) & $x$ & 
$N_{\scriptsize{\textrm{data}}}$ &
$N_{\scriptsize{\textrm{bg}}}$ &
$\delta_{\scriptsize{\textrm{unc}}}(\%)$ &
$\delta_{1}(\%)$ & 
$\delta_{2}(\%)$ & 
$\delta_{3}(\%)$ & 
$\tilde \sigma $ \\
    \hline
     280 & 0.008 & 
   26 &  2.5 & 
$ ~^{+   2.8}_{-  2.8} $& 
$ ~^{+   4.4}_{-  2.5} $& 
$ ~^{+   0.3}_{+  2.1} $& 
$ ~^{-   5.3}_{+  5.3} $& 
$(1.47\pm0.28^{+0.11}_{-0.09}) \cdot 10^{  0}$ \\ 
     280 & 0.015 & 
   49 &  5.3 & 
$ ~^{+   1.9}_{-  3.0} $& 
$ ~^{+   1.4}_{-  2.2} $& 
$ ~^{-   0.6}_{-  0.1} $& 
$ ~^{-    10}_{+   10} $& 
$(1.08\pm0.15^{+0.11}_{-0.12}) \cdot 10^{  0}$ \\ 
     280 & 0.032 & 
   55 &  3.3 & 
$ ~^{+   1.5}_{-  1.9} $& 
$ ~^{+   2.1}_{+  0.2} $& 
$ ~^{+   1.6}_{-  0.6} $& 
$ ~^{-   3.1}_{+  3.1} $& 
$(8.52\pm1.12^{+0.37}_{-0.31}) \cdot 10^{ -1}$ \\ 
     280 & 0.068 & 
   24 &  2.8 & 
$ ~^{+   0.8}_{-  5.1} $& 
$ ~^{+   0.6}_{-  4.0} $& 
$ ~^{-   0.7}_{-  1.7} $& 
$ ~^{-   1.0}_{+  1.0} $& 
$(4.14\pm0.78^{+0.06}_{-0.28}) \cdot 10^{ -1}$ \\ 
     530 & 0.015 & 
   52 &  0.5 & 
$ ~^{+   1.0}_{-  1.0} $& 
$ ~^{+   2.1}_{-  0.7} $& 
$ ~^{-   0.9}_{+  0.3} $& 
$ ~^{-   7.6}_{+  7.6} $& 
$(8.39\pm1.18^{+0.67}_{-0.65}) \cdot 10^{ -1}$ \\ 
     530 & 0.032 & 
   57 &  0.7 & 
$ ~^{+   1.0}_{-  0.8} $& 
$ ~^{+   0.5}_{-  1.3} $& 
$ ~^{-   0.2}_{+  0.9} $& 
$ ~^{+   3.7}_{-  3.7} $& 
$(6.15\pm0.82^{+0.24}_{-0.25}) \cdot 10^{ -1}$ \\ 
     530 & 0.068 & 
   59 &  0.6 & 
$ ~^{+   1.8}_{-  1.3} $& 
$ ~^{+   1.6}_{+  0.5} $& 
$ ~^{+   2.1}_{+  0.9} $& 
$ ~^{-   2.4}_{+  2.4} $& 
$(6.28\pm0.83^{+0.26}_{-0.17}) \cdot 10^{ -1}$ \\ 
     530 & 0.130 & 
   25 &  0.2 & 
$ ~^{+   7.2}_{-  1.1} $& 
$ ~^{+   4.2}_{+  1.2} $& 
$ ~^{+   4.1}_{+  2.5} $& 
$ ~^{+    10}_{-   10} $& 
$(4.40\pm0.89^{+0.61}_{-0.44}) \cdot 10^{ -1}$ \\ 
     950 & 0.015 & 
   52 &  2.0 & 
$ ~^{+   2.0}_{-  1.6} $& 
$ ~^{+   1.1}_{+  0.6} $& 
$ ~^{+   1.5}_{-  0.9} $& 
$ ~^{-   8.5}_{+  8.5} $& 
$(6.85\pm0.94^{+0.61}_{-0.59}) \cdot 10^{ -1}$ \\ 
     950 & 0.032 & 
  102 &  0.8 & 
$ ~^{+   0.5}_{-  0.6} $& 
$ ~^{+   1.5}_{-  0.7} $& 
$ ~^{-   0.4}_{+  0.5} $& 
$ ~^{-   5.4}_{+  5.4} $& 
$(6.97\pm0.70^{+0.40}_{-0.39}) \cdot 10^{ -1}$ \\ 
     950 & 0.068 & 
   84 &  0.8 & 
$ ~^{+   0.5}_{-  1.6} $& 
$ ~^{-   0.4}_{-  0.9} $& 
$ ~^{-   1.0}_{-  0.2} $& 
$ ~^{+   2.0}_{-  2.0} $& 
$(5.63\pm0.62^{+0.11}_{-0.16}) \cdot 10^{ -1}$ \\ 
     950 & 0.130 & 
   48 &  0.2 & 
$ ~^{+   0.5}_{-  4.0} $& 
$ ~^{+   0.6}_{-  3.8} $& 
$ ~^{-   0.4}_{-  2.7} $& 
$ ~^{+   1.4}_{-  1.4} $& 
$(4.86\pm0.71^{+0.08}_{-0.31}) \cdot 10^{ -1}$ \\ 
     950 & 0.240 & 
   20 &  0.3 & 
$ ~^{+   5.1}_{-  1.1} $& 
$ ~^{+   0.9}_{-  0.7} $& 
$ ~^{-   4.5}_{-  0.4} $& 
$ ~^{-   4.4}_{+  4.4} $& 
$(2.87\pm0.64^{+0.20}_{-0.19}) \cdot 10^{ -1}$ \\ 
    1700 & 0.032 & 
  105 &  0.9 & 
$ ~^{+   0.9}_{-  0.8} $& 
$ ~^{-   1.0}_{-  0.6} $& 
$ ~^{+   0.4}_{-  0.0} $& 
$ ~^{+   3.2}_{-  3.2} $& 
$(5.55\pm0.55^{+0.19}_{-0.19}) \cdot 10^{ -1}$ \\ 
    1700 & 0.068 & 
  105 &  0.2 & 
$ ~^{+   1.2}_{-  0.5} $& 
$ ~^{+   1.1}_{+  0.4} $& 
$ ~^{-   0.1}_{+  0.5} $& 
$ ~^{-   1.0}_{+  1.0} $& 
$(4.84\pm0.48^{+0.10}_{-0.05}) \cdot 10^{ -1}$ \\ 
    1700 & 0.130 & 
   57 &  0.2 & 
$ ~^{+   1.3}_{-  0.6} $& 
$ ~^{+   0.4}_{+  0.9} $& 
$ ~^{+   1.5}_{-  0.2} $& 
$ ~^{+   3.3}_{-  3.3} $& 
$(3.71\pm0.50^{+0.15}_{-0.13}) \cdot 10^{ -1}$ \\ 
    1700 & 0.240 & 
   39 &  0.1 & 
$ ~^{+   1.3}_{-  3.4} $& 
$ ~^{-   0.4}_{+  0.4} $& 
$ ~^{-   0.5}_{+  0.2} $& 
$ ~^{-   3.6}_{+  3.6} $& 
$(2.73\pm0.44^{+0.10}_{-0.14}) \cdot 10^{ -1}$ \\ 
    3000 & 0.068 & 
   97 &  0.3 & 
$ ~^{+   0.9}_{-  1.6} $& 
$ ~^{-   2.2}_{+  2.2} $& 
$ ~^{+   2.1}_{-  1.3} $& 
$ ~^{-   1.1}_{+  1.1} $& 
$(3.71\pm0.39^{+0.13}_{-0.12}) \cdot 10^{ -1}$ \\ 
    3000 & 0.130 & 
   55 &  0.1 & 
$ ~^{+   1.0}_{-  0.5} $& 
$ ~^{-   2.0}_{+  1.6} $& 
$ ~^{-   1.4}_{+  1.9} $& 
$ ~^{-   3.2}_{+  3.2} $& 
$(2.73\pm0.37^{+0.12}_{-0.11}) \cdot 10^{ -1}$ \\ 
    3000 & 0.240 & 
   44 &  0.1 & 
$ ~^{+   2.3}_{-  0.5} $& 
$ ~^{-   2.1}_{+  2.1} $& 
$ ~^{-   1.3}_{+  1.8} $& 
$ ~^{-   3.0}_{+  3.0} $& 
$(2.17\pm0.33^{+0.10}_{-0.08}) \cdot 10^{ -1}$ \\ 
    3000 & 0.420 & 
    7 &  0.0 & 
$ ~^{+   3.5}_{-  1.3} $& 
$ ~^{-   0.4}_{+  1.4} $& 
$ ~^{-   2.3}_{+  1.7} $& 
$ ~^{-   3.1}_{+  3.1} $& 
$(4.24^{+2.29}_{-1.57}~^{+0.22}_{-0.17}) \cdot 10^{ -2}$ \\ 
    5300 & 0.068 & 
   49 &  0.1 & 
$ ~^{+   1.0}_{-  1.4} $& 
$ ~^{-   2.7}_{+  2.3} $& 
$ ~^{+   1.3}_{-  2.5} $& 
$ ~^{+   2.3}_{-  2.3} $& 
$(2.28\pm0.33^{+0.08}_{-0.10}) \cdot 10^{ -1}$ \\ 
    5300 & 0.130 & 
   45 &  0.0 & 
$ ~^{+   0.8}_{-  0.8} $& 
$ ~^{-   1.4}_{+  2.5} $& 
$ ~^{+   1.0}_{+  0.1} $& 
$ ~^{-   1.2}_{+  1.2} $& 
$(1.96\pm0.29^{+0.06}_{-0.04}) \cdot 10^{ -1}$ \\ 
    5300 & 0.240 & 
   41 &  0.0 & 
$ ~^{+   0.9}_{-  0.7} $& 
$ ~^{-   2.4}_{+  3.7} $& 
$ ~^{-   2.0}_{+  4.4} $& 
$ ~^{-   0.7}_{+  0.7} $& 
$(1.80\pm0.28^{+0.11}_{-0.06}) \cdot 10^{ -1}$ \\ 
    5300 & 0.420 & 
   20 &  0.0 & 
$ ~^{+   1.1}_{-  3.9} $& 
$ ~^{-   2.7}_{+  3.8} $& 
$ ~^{-   2.4}_{+  4.2} $& 
$ ~^{+   3.6}_{-  3.6} $& 
$(9.85\pm2.22^{+0.67}_{-0.63}) \cdot 10^{ -2}$ \\ 
    9500 & 0.130 & 
   21 &  0.1 & 
$ ~^{+   3.5}_{-  2.4} $& 
$ ~^{-   6.7}_{+  5.8} $& 
$ ~^{+   5.2}_{-  5.7} $& 
$ ~^{+   3.9}_{-  3.9} $& 
$(9.99\pm2.18^{+0.94}_{-0.99}) \cdot 10^{ -2}$ \\ 
    9500 & 0.240 & 
   22 &  0.0 & 
$ ~^{+   1.0}_{-  2.2} $& 
$ ~^{-   4.2}_{+  3.6} $& 
$ ~^{-   3.5}_{+  1.3} $& 
$ ~^{+   3.6}_{-  3.6} $& 
$(1.03\pm0.22^{+0.06}_{-0.07}) \cdot 10^{ -1}$ \\ 
    9500 & 0.420 & 
    8 &  0.0 & 
$ ~^{+   3.4}_{-  0.6} $& 
$ ~^{-   7.6}_{+  6.1} $& 
$ ~^{-   6.2}_{+  5.7} $& 
$ ~^{+   0.3}_{-  0.3} $& 
$(3.78^{+1.86}_{-1.31}~^{+0.34}_{-0.37}) \cdot 10^{ -2}$ \\ 
   17000 & 0.240 & 
    3 &  0.0 & 
$ ~^{+   3.9}_{-  2.8} $& 
$ ~^{-   8.8}_{+   11} $& 
$ ~^{+   6.1}_{-  5.0} $& 
$ ~^{+   6.5}_{-  6.5} $& 
$(1.55^{+1.49}_{-0.84}~^{+0.23}_{-0.19}) \cdot 10^{ -2}$ \\ 
   17000 & 0.420 & 
    6 &  0.0 & 
$ ~^{+   0.7}_{-  3.8} $& 
$ ~^{-   7.0}_{+  7.8} $& 
$ ~^{-   6.5}_{+  5.6} $& 
$ ~^{+   2.8}_{-  2.8} $& 
$(3.19^{+1.91}_{-1.27}~^{+0.32}_{-0.34}) \cdot 10^{ -2}$ \\ 
    \hline
    \end{tabular}
\end{center}
\caption{Values of the reduced cross section. The following quantities are given: the values of $Q^2$ and $x$ at which the cross section is quoted; the number of data events; the number of expected background events; the uncorrelated systematic uncertainty; those systematic uncertainties with correlations between cross-section bins $\delta_1$, $\delta_2$ and $\delta_3$ defined in Section~\ref{s:systerr} and the measured cross section, with statistical and total systematic uncertainties. The uncertainty on the measured luminosity of $2.5\%$ is not included in the total systematic uncertainty.}
\label{t:double}
\end{table}


\newpage

\end{document}